\documentclass[pre,aps,twocolumn,superscriptaddress]{revtex4-2}
\usepackage{amssymb}
\usepackage{epsfig}
\usepackage{amsmath}
\usepackage{subfigure}
\usepackage{graphicx}
\usepackage{textcomp}
\usepackage{url}
\usepackage{float}
\usepackage{color}
\usepackage{cases}
\usepackage[normalem]{ulem}

\usepackage[titletoc,title]{appendix}

\usepackage[colorlinks=true, urlcolor=blue, anchorcolor=blue, citecolor=blue,filecolor=blue,linkcolor=blue,menucolor=blue]{hyperref}

\usepackage{color}

\begin{document}
\title{Probabilities of moderately  atypical fluctuations of the size of a swarm of Brownian Bees}

\author{Pavel Sasorov}
\email{pavel.sasorov@gmail.com}
\affiliation{Institute of Physics CAS, ELI Beamlines, 182 21 Prague, Czech Republic}
\author{Arcady Vilenkin}
\email{vilenkin@mail.huji.ac.il}
\affiliation{Racah Institute of Physics, Hebrew University of
Jerusalem, Jerusalem 91904, Israel}

\author{Naftali R. Smith}
\email{naftalismith@gmail.com}
\affiliation{Department of Solar Energy and Environmental Physics, Blaustein Institutes for Desert Research,
Ben-Gurion University of the Negev, Sede Boqer, 8499000, Israel}

\begin{abstract}
The ``Brownian bees'' model describes an ensemble of $N=$~const independent branching Brownian particles. The conservation of $N$ is provided by a modified branching process. When a particle branches into two particles, the particle which is farthest from the origin is eliminated simultaneously. The spatial density of the particles is governed by the solution of a free boundary problem for a reaction-diffusion equation in the limit of $N \gg 1$. At long times, the particle density approaches a spherically symmetric steady state solution with a compact support of radius $\bar{\ell}_0$. However, at finite $N$, the radius of this support, $L$,  fluctuates. The variance of these fluctuations appears to exhibit a logarithmic anomaly [Siboni {\em et al}., Phys. Rev. E. {\bf104}, 054131 (2021)]. It is proportional to $N^{-1}\ln N$ at $N\to\infty$. We investigate here the tails of the probability density function (PDF), $P(L)$, of the swarm radius, when the absolute value of the radius fluctuation $\Delta L=L-\bar{\ell}_0$ is sufficiently larger than the typical fluctuations' scale determined by the variance. For negative deviations the PDF can be obtained in the framework of the optimal fluctuation method (OFM). This part  of the PDF displays the scaling behavior: $\ln P\propto - N \Delta L^2\, \ln^{-1}(\Delta L^{-2})$, demonstrating a logarithmic anomaly at small negative $\Delta L$. For the opposite sign of the fluctuation,  $\Delta L > 0$, the PDF can be obtained with an approximation of a single particle, running away. We find that $\ln P \propto -N^{1/2}\Delta L$. We consider in this paper only the case, when $|\Delta L|$ is much less than the typical radius of the swarm at $N\gg 1$.
\end{abstract}

\maketitle

\nopagebreak

\section{Introduction}
\label{intro}

We continue in this paper investigations of a model of nonequilibrium statistical physics, which is known under the name `Brownian bees'~\cite{bees1,bees2,MS21,Si21}. This model combines two important fields of statistical physics: branching Brownian motion (BBM) and nonequilibrium steady states (NESSs). BBM includes two process: Brownian motion together with a branching  process. Growing ensembles described by this model have been investigated for a long time. See for example Refs.~\cite{McKean,Bramson} and more recent Refs.~\cite{BD09,Mue14,Ram15,DMS}.
In its turn, ensembles of reacting and diffusing particles, representing NESSs, are important for desription of many natural systems. Their investigations occupy  a very distinguishable area in nonequilibrium statistical mechanics \cite{JL1993,JL2004,Bodineau2010,Hurtado2013,M2015}.

The ``Brownian bees'' model represents a system whose dynamics are irreversible in time,  based on the branching Brownian dynamics of $N$ particles (bees). Conservation of their total number is provided by removing the bee  that is farthest from the origin at the moment of any branching. The origin of the swarm is assumed to be immobile. The removing causes nonlocal  interaction between bees and destroys time reversibility of the system even in its steady state. Choosing proper units for time and distance we may set that the diffusion coefficient for the Brownian motion and rate of branching of each bee are equal to 1. Most of this paper is devoted to the 1-dimensional case at $N\to\infty$.

It has been shown~\cite{bees1} that at any finite time $t$ a coarse-grained density distribution $u(x,t)$ of the bees, normalized by $N$,  obeys the following mean field theory at $N\to\infty$:
\begin{eqnarray}
  &&\partial_t u (x,t) = \partial_x^2 u(x,t)+u(x,t)\,,\quad |x|\leq \bar{\ell}(t)\,, \label{INT10}\\
	  &&u(x,t)=0\,, \quad |x|> \bar{\ell}(t)\,,
  \label{LAN110} \\
	  &&\int_{-\bar{\ell}(t)}^{\bar{\ell}(t)}u(x,t)\, dx =1\,.
  \label{LAN120}
\end{eqnarray}
As one can see, the compact support of $u(x,t)$, at all finite $t>0$, is centered at the origin. Effectively, there are two absorbing walls, at $x=\pm \bar{\ell}(t)$, which move in synchrony so as to keep the number of particles constant at all times. The results of Ref.~\cite{bees1} argue that fluctuations of the coarse-grained density around $u(x,t)$ tend to 0 at $N\to\infty$.

It has been proved also~\cite{bees2} that the general solution of the system~(\ref{INT10})-(\ref{LAN120}) tends at $t\to\infty$ to the following steady state:
\begin{equation}\label{Ux}
U\left(x\right)=\begin{cases}
{\displaystyle \frac{1}{2}}\cos x\,, & |x|\le\bar{\ell}_{0},\\[4mm]
0\,, & |x|>\bar{\ell}_{0},
\end{cases}
\end{equation}
where $\bar{\ell}_0=\pi/2$, and $\bar{\ell}(t)\to \bar{\ell}_0$ at $t\to\infty$.

We consider in this paper small relative fluctuations of the swarm radius $L=\max{|x|}$ in the steady state described by Eq.~(\ref{Ux}). Consideration of this problem started in Refs.~\cite{MS21,Si21}. Monte-Carlo simulations of the initial microscopic model and analytic investigation of a Langevin equation that describes typical fluctuations in this model gave the following expression~\cite{Si21} for the variance of $L$ at the steady state at $N\to\infty$:
\begin{equation}\label{L140}
\textrm{var}\,L \simeq \frac{2}{\pi} \frac{\ln N}{N}\,.
\end{equation}
This result was obtained in Ref.~\cite{Si21} by linearization of the Langevin equation; and existence of the logarithmic anomaly indicates that it was a truncation of formally divergent analytic expression for $\textrm{var}\,L$. This anomaly originates from the fact that fluctuations of the particle density at all spatial scales give contributions that are of the same order of magnitude to the fluctuations of $L$.

The ``Brownian bees'' model belongs to a broader class of $N$-particle branching Brownian models with selection (NBBM). It was introduced initially in Refs.~\cite{BDMM2006,BDMM2007}. A lot of works investigating this class of models are cited in Ref.~\cite{Si21}. It is interesting that many NBBM systems expose logarithmic anomalies in the statistical behavior of the edge particles. This is an additional motivation for studies of the Brownian bees model.

In this paper, we present results of our investigation of the tails of the probability density function (PDF) of instantaneous values of $L$ at the steady state, $P(L)$. We consider sufficiently moderately-large fluctuations, $\Delta L=L-\ell_0$, which on the one hand are much larger than the typical fluctuations' scale, $|\Delta L|\gg \sigma(L)=\sqrt{\textrm{var}\,L}$, but on the other hand, are relatively small fluctuations in the sense that $|\Delta L|\ll1$.

Let us briefly summarize our main findings, while describing the structure of the rest of the paper.
For such negative $\Delta L$, the fluctuations involve many particles. So,  they can be considered in the framework of the optimal fluctuation method (OFM). The latter, known also under the other names (the instanton method, the weak noise theory, and the macroscopic fluctuation theory), considers a single `trajectory' of coarse grained density history, giving maximal contribution to the probability~\cite{EK,MS,MSK,MFT,JL1993,JL2004,MSFKPP,MVS}. It is briefly recalled in Sec.~\ref{GE}.  Applying the OFM to the present problem in Sec.~\ref{OFM}, we obtain at $N\to\infty$:
\begin{equation}\label{I040}
-\ln P(L) =  \frac{\pi}{4} N \frac{\Delta L^2}{|\ln \Delta L^2|}+\dots\,.
\end{equation}
This result is obtained by combining analytical and numerical methods.
Existence of the logarithm in this asymptotic expression means that it can hardly be obtained by regular perturbation methods at $-\Delta L\ll 1$. However, the non-analytic structure of this expression at $-\Delta L\to 0$ provides a smooth matching of this result with the Gaussian distribution that describes typical fluctuations with mean $\bar{\ell}_0$ and variance~(\ref{L140}).

\begin{figure}[ht]
  \centering
  \includegraphics[width=8cm]{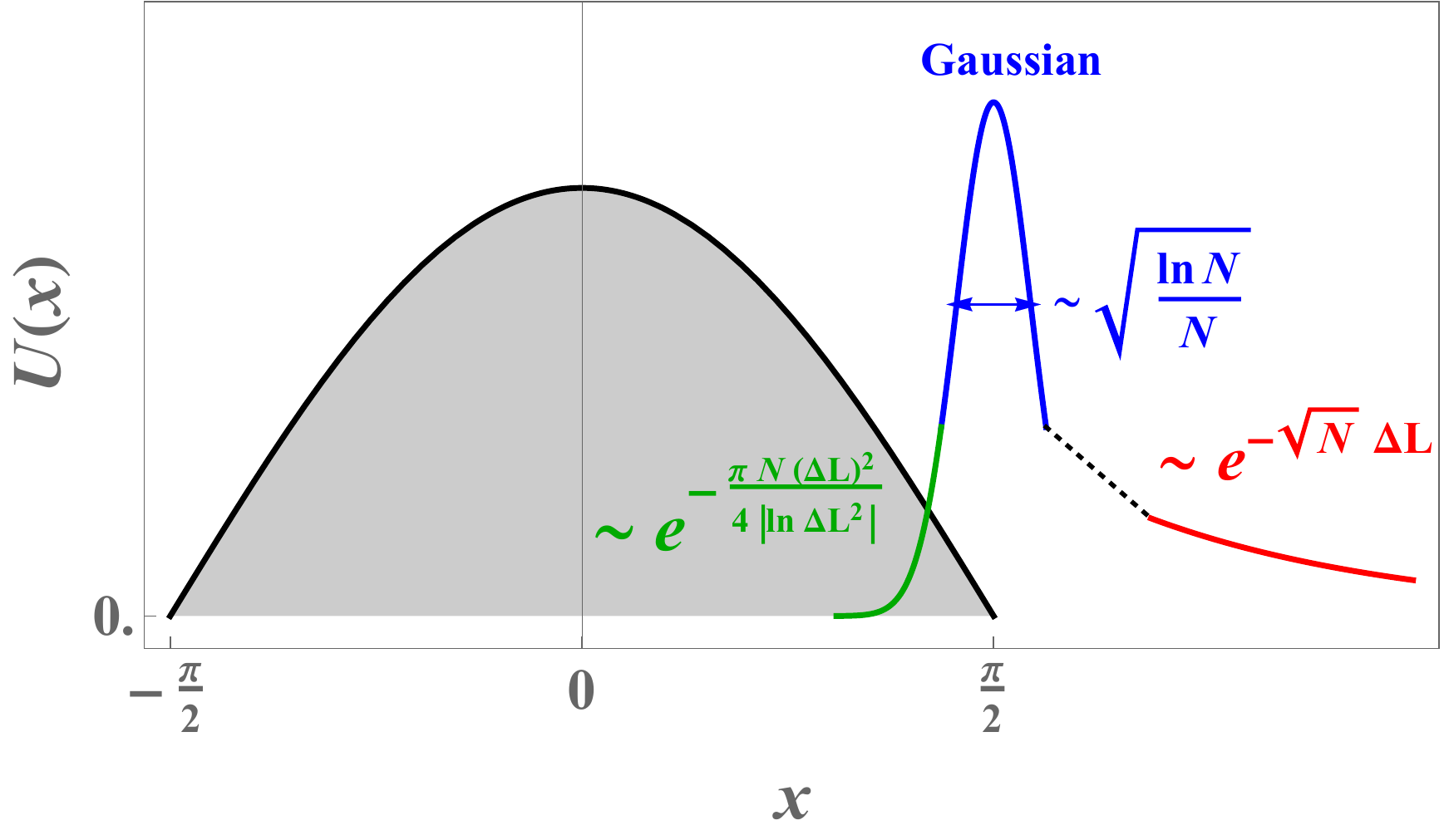}
  \caption{A schematic plot of the distribution of the swarm radius, $P(L)$. The peak of the distribution is at $x=\pi/2$ which is the edge of the mean-field swarm density $U(x)$. Typical fluctuations are Gaussian with variance \eqref{L140} \cite{Si21}, while the two, very asymmetric large deviation tails of $P(L)$ are described by Eqs.~\eqref{I040} and \eqref{I050}. The dotted line corresponds to a conjectured crossover regime between the typical fluctuations and the atypical positive fluctuations regime.} 
  \label{figPLSchematic}
\end{figure}

Atypically large positive fluctuations of $L$ turn out to be dominated by the dynamical behavior of the single farthest particle. Analogous approaches were applied in many other problems of extreme value statistics \cite{MPS20}. This  approach, applied in Sec.~\ref{SP} for the present problem, gives at $N\gg1$:
\begin{equation}\label{I050}
-\ln P(L) = \sqrt{N}\, \Delta L+\dots\,.
\end{equation}
The result in this regime does not match smoothly with the typical-fluctuations, Gaussian regime. We expect there to be a crossover between the two regimes which we do not attempt to analyze in the present work.
Our main results for the distribution $P(L)$ are plotted schematically in Fig.~\ref{figPLSchematic}.
Finally, Sec.~\ref{summ} is devoted to conclusions and discussions, including generalizations of our results to higher dimensions.

\section{Optimal fluctuation method: governing equations}
\label{GE}

The OFM employs an  idea that probability of transition between two states of a stochastic system is dominated by the probability of an ``optimal'' (most likely) trajectory. In our case this trajectory is given by $q(x,t)$: A coarse grained normalized density of the fluctuating swarm of Brownian Bees. The designation $u(x,t)$ we reserve for particular trajectories $q(x,t)$, obeying the mean field system of equations~(\ref{INT10})-(\ref{LAN120}) (without fluctuations). The most probable trajectory is defined by minimization of a functional of $q(x,t)$. Hence, this problem corresponds to the investigation of some classical field theory that can be recast as a Hamiltonian field theory. It involves, in addition to a generalized `coordinate' $q(x,t)$ a generalized `momentum' field $p(x,t)$. Before introducing the classical field theory corresponding to our present system, we may say that typically the OFM gives relevant evaluation of the initial probability for atypical fluctuations when they involved many particles of an initial many-particle system.

The derivation of the OFM for the Brownian bees models was considered in detail together with references of previous publications in Ref.~\cite{MS21}. The process providing the conservation of total number of bees, $N$, in the swarm is introduced in the OFM system as a constrain and as boundary conditions at $|x|=L(t)$, whereas the fields $p(x,t)$ and $q(x,t)$ are defined only at $x\in[-L(t), L(t)]$. The  density of the unconstrained Hamiltonian is defined as:
\begin{equation}
\label{H0am}
\mathcal{H}_0(q,p) =(e^p-1) q -(\partial_x q) (\partial_x p)
+q \left(\partial_x p\right)^2\, ,
\end{equation}
whereas the unconstrained Hamiltonian is equal to
\begin{equation}
\label{H0}
H_0[q(x),p(x)] =\int\limits_{-L(t)}^{L(t)} \mathcal{H}_0(q,p)\, dx\, .
\end{equation}
The constraint coming from conservation of the total number of bees
\begin{equation}\label{conservq}
 \int\limits_{-L(t)}^{L(t)}  q(x,t) \,dx = 1 \quad \text{for any } t
\end{equation}
is introduced by means of a Lagrangian multiplier $\lambda(t)$, so that constrained Hamiltonian becomes
\begin{equation}
\label{H010}
H[q(x),p(x),\lambda(t)] =H_0[q(x),p(x)]+\lambda(t) \!\int\limits_{-L(t)}^{L(t)}  q(x,t) \,dx\, .
\end{equation}
It depends, in general, explicitly on time $t$. The  density of the constrained Hamiltonian is
\begin{equation}
\label{H020}
\mathcal{H}(q,p)=\mathcal{H}_0(q,p) +\lambda(t) q\, .
\end{equation}
The fields $q(x,t)$ and $p(x,t)$ have support at $|x|<L(t)$, whereas the boundary conditions at the absorbing wall, $|x|=L(t)$, are
\begin{equation}
\label{H030}
q(|x|=L(t),t)=p(|x|=L(t),t)=0\, .
\end{equation}
Considering an instantaneous fluctuations at $t=0$ over the steady state $U(x)$ we may demand that at $t\to-\infty$ the system should be in the steady state. This condition generates the following initial condition:
\begin{equation}
\label{H040}
q(x,t\to-\infty)=U(x)\, ;\quad
p(x,t\to-\infty)=0\, ,
\end{equation}
and hence $L(t\to-\infty)=\bar{\ell}_0$.
Trajectories of this Hamilton system, determined by the Hamilton equations
\begin{eqnarray}
\!\!\!\!\!  \partial_t q &=& \frac{\delta H}{\delta p} = q e^p +\nabla \cdot \left(\nabla q-2 q \nabla p\right),
  \label{H062} \\
\!\!\!\!\!  \partial_t p &=& -\frac{\delta H}{\delta q} = -\left(e^p-1\right) - \nabla^2 p- (\nabla p)^2 -\lambda(t),
  \label{H064}
\end{eqnarray}
maximize locally the probability density, $\mathcal{P}[q(x,t)]$, at the trajectory $q(x,t)$ which is determined in the OFM framework by the action  $S[q(x,t)]$ of an unconstrained mechanical system:
\begin{equation}
\label{H050}
-N^{-1}\, \ln \mathcal{P}[q(x,t)]=S[q(x,t)] \, ,
\end{equation}
where the action functional $S[q(x,t)]$ on an arbitrary trajectory   $q(x,t)$ per particle is defined as
\begin{equation}
\label{H060}
S[q(x,t)] = \int\limits_{-\infty}^0 dt\int\limits_{-L(t)}^{L(t)}\left[p\partial_t q-\mathcal{H}_0\right]\, dx\, .
\end{equation}
Here, the momentum field $p$ should obey Eq.~(\ref{H062}) as usual, and the boundary condition~(\ref{H030}).
Minimization of the action with respect to small variations $\delta q(x,t)$ of the trajectory $q(x,t)$ at $t<0$ gives the 2nd Hamilton equation~(\ref{H064}). This minimization is necessary because  the OFM implies the following evaluation of the probability that $L<\ell$ for $\ell<\bar{\ell}_0$:
\begin{equation}
\label{HH062}
-N^{-1} \ln\mathbb{P}\mbox{rob}\, (L<\ell) \simeq \min\limits_{L(0)=\ell} S[q(x,t)]\, .
\end{equation}
It corresponds to Eq.~(\ref{H050}) and to the following expression for the PDF, $P(L)$:
\begin{equation}
\label{HH064}
-N^{-1} \ln P(L) \simeq \min\limits_{L(0)=L} S[q(x,t)]\, ,
\end{equation}
where the minimization is over all possible trajectories $q(x,t)$ obeying the constraints.

The minimization, entering Eq.~(\ref{HH064}), means in particular minimization over the final density $q(x,0)$ of the particles inside the interval $|x|<L$ at $t=0$. Requiring the variation of $S[q(x,t)]$ over $q(x,0)$ to vanish, conditioned on $\int q(x,0)\, dx =1$,  that is equivalent to
$\int\delta q(x,0)\, dx =0$, gives a boundary condition at $t=0$. We may follow Ref.~\cite{DG} to get analogously this boundary condition for the present problem. Consider two solutions of Eqs.~(\ref{H062}) and (\ref{H064}) that are close to each other, $q(x,t)$ and $q(x,t)+\delta q(x,t)$, obeying the boundary conditions~(\ref{H040}) at $t=-\infty$. Then we may write for the variation $\delta S$:
$$
\delta S=\int\limits_{-\infty}^0 dt\int\limits_{-L(t)}^{L(t)}\left[\delta p\partial_t q+p\partial_t \delta q-\frac{\partial\mathcal{H}_0}{\partial q}\delta q-\frac{\partial\mathcal{H}_0}{\partial p}\delta p\right] dx.
$$
Using Eqs.~(\ref{H062}) and (\ref{H064}), we obtain:
$$
\delta S=\int\limits_{-\infty}^0 dt\int\limits_{-L(t)}^{L(t)}\left[\partial_t (p\delta q)+\lambda(t)\delta q\right]\, dx.
$$
Applying the boundary condition~(\ref{H040}) and the condition~(\ref{conservq}), we obtain:
$$
\delta S=\int\limits_{-L(0)}^{L(0)}p(x,0)\delta q(x,0)\, dx.
$$
For the optimal $q(x,t)$ the variation $\delta S$ should vanish. Combining this requirement with the previous equation and with Eq.~(\ref{conservq}), we obtain that $\partial_xp(x,0)=0$ and hence
\begin{equation}
\label{H080}
p(x,0)=\Lambda=\mbox{const}\quad \text{for }|x|<L(0)\, .
\end{equation}
This relationship gives the last boundary condition in time $t$ for our problem. 

The problem~(\ref{H062}), (\ref{H064}), (\ref{H030}), (\ref{H040}), (\ref{H080}) and~(\ref{conservq}) contains one constant $\Lambda$, and two unknown yet functions $\lambda(t)$ and $L(t)$. The latter one defines also the constant $L(0)$. When the function $\lambda(t)$ is known and tends to 0 sufficiently fast at $t\to-\infty$, then the condition of solvability of the system~(\ref{H062}), (\ref{H064}), (\ref{H030}), (\ref{H040}), (\ref{H080}) and~(\ref{conservq}) determines $\Lambda$ and $L(t)$ (and hence $L(0)$). As a result, solution of the system~(\ref{H062}), (\ref{H064}), (\ref{H030}), (\ref{H040}), (\ref{H080}) and~(\ref{conservq}) demonstrates a functional degree of freedom that is determined by the choice of $\lambda(t)$. Thus our action $S$ is actually  a functional of $\lambda(t)$: $S=S[\lambda(t)]$. An equation that follows from the condition of vanishing of variational derivative: $\delta S[\lambda(t)] / \delta \lambda  =0$ under constrain that $L(0)=L$ looks as a very cumbersome and almost useless. We will try in Sec.~\ref{OFM} to find an approximation to an optimal $\lambda(t)$ at $\Delta L=L(0)-\bar{\ell}_0\to 0$ with another approach, that will give a leading order of the optimal action $S$ in this limit. Note, that we may not distinguish $L$ and $L_0=L(0)$ in the frame of the OFM.
%
For the {\em locally} optimal at $t<0$ trajectories, the general expression~(\ref{H060}) for the action becomes simpler:
\begin{equation}\label{H090}
S=\int\limits_{-\infty}^0 dt \int\limits_{-L(t)}^{L(t)} dx\, \left[q \left(p e^p -e^p+1\right) + q (\partial_x p)^2 \right]\,.
\end{equation}
Integrating the 1st term in Eq.~(\ref{H060}) by parts, and using Eq.~(\ref{H064}), we obtain even simpler expression for the action:
\begin{equation}\label{H100}
S=\Lambda+\int\limits_{-\infty}^0 \lambda(t) dt \,.
\end{equation}
However, the latter expression is not so suitable for computer simulations at $\Delta L\to 0$, because as we will see, each term in Eq.~(\ref{H100}) behaves as ${\cal O}(1)$ in this limit, whereas $S\to 0$, as it can be seen from Eq.~(\ref{I040}). Both terms in Eq.~(\ref{H090}) are positive-definite. This property is much more suitable for numerical applications.

The OFM described briefly above may give relevant estimation for the PDF $P(L)$, when the rare enough fluctuation touches a lot of particles of the system. this situation takes place for negative $\Delta L$, when its absolute value is significantly larger than $\sqrt{\mbox{var}\,L}$. More exact criteria will be considered in Sec.~\ref{OFM}, when we will obtain our asymptotic expression for $S(L)$. For sufficiently large positive $\Delta L$ the situation is quite different, and our evaluation of the PDF $P(L)$ for this case cannot be obtained in the frame of the OFM, because of a completely different scaling with $N$.

\medskip

\section{Negative atypical  fluctuations of $L$}
\label{OFM}

We try in this section to obtain a solution of the mathematical problem we set in previous Sec.~\ref{GE} for sufficiently small negative $\Delta L$, $|\Delta L| \ll 1$. Our main obstacle to do this is how to determine $\lambda(t)$ that minimizes the action functional $S[\lambda(t)]$. However, we are able to obtain a solution to the problem within a quite reasonable class of functions $\lambda(t)$, that may give negative $\Delta L$ tending to 0. We obtain such solutions numerically as well as analytically. The latter one concerns only a leading order of the solution at $\Delta L\to 0$. Such an approach would appear to give only an upper boundary for $S(L)$. However, our final results show that $S(L)/\Delta L^2\to 0$ at $\Delta L\to 0^-$. Such behavior is only possible for quite specific forms of $\lambda(t)$, so that an optimal $\lambda(t)$ is determined almost uniquely at $(-t)\ll1$ as well as a leading term of $S(L)$ asymptotics at  $\Delta L\to 0^-$.

We introduce in Sec.~\ref{ps} a one parametric set of particular $\lambda(t)$ and investigate analytically and numerically such OFM solutions including calculation of the action at $\Delta L\to 0^-$. Then we explain in Sec.~\ref{GR} that the upper bound for the action obtained in this way at $\Delta L\to 0^-$ has the same leading order of the true action $S(L)$, calculated along optimal trajectories at $\Delta L\to 0^-$.

\subsection{A particular choice for $\lambda(t)$}
\label{ps}

We consider in this section the following choice of one parametric set for $\lambda(t)$.
\begin{widetext}
\begin{equation}\label{lam20}
\lambda(-\infty<t<0)=- \frac{\sqrt{t_\lambda}}{4}\times\left\{
\begin{array}{ll}
64\times 3^{-3/2}\times e^{3/2+8(t-t_\lambda)}&\mbox{~~~for~~~~~}t<-3/16+t_\lambda , \\[3mm]
(t_\lambda-t)^{-3/2}&\mbox{~~~for~~~}-3/16+t_\lambda<t<0 .
\end{array}
\right.
\end{equation}
\end{widetext}
This set of $\lambda(t)$ has the single positive parameter $t_\lambda$: $0<t_\lambda\ll1$. Solution of the problem~(\ref{H062}), (\ref{H064}), (\ref{H030}), (\ref{H040}), (\ref{H080}) and~(\ref{conservq}) with such $\lambda(t)$ gives in particular the value of $L(0)$ that depends on $t_\lambda$. We will see that $L(0)$ depends monotonically on $t_\lambda$ at small $t_\lambda$, and $L(0)\to \bar{\ell}_0^-$ at $t_\lambda$ tending to 0. We will see also that leading order of the action~(\ref{H090}) along the trajectories defined by such $\lambda(t)$ is determined by times $t_\lambda\ll (-t)\ll 1$ and corresponds to the expression:
\begin{equation}\label{H110}
S\left(L(0)\right) =\frac{\pi}{4}\frac{\Delta L^2}{|\ln \Delta L^2|}+\dots\,,
\end{equation}
whereas the parts of the trajectories on the time intervals $0<(-t)\lesssim t_\lambda$ and $1\lesssim(-t) < \infty$ contribute only to the subleading term in Eq.~(\ref{H110}) at $\Delta L\to 0^-$. We will see also in Sec.~\ref{GR} that introducing of a constant multiplier in Eq.~(\ref{lam20}) of the order of ${\cal O}(1)$ at $\Delta L \to 0^-$ does not change the leading order in Eq.~(\ref{H110}) and influences only on the subleading order.

The most important part of the trial function $\lambda(t)$ corresponds to the second line in Eq.~(\ref{lam20}). Its possible form for optimization of the action will be considered in detail in Sec.~\ref{GR}. In this section we treat it as a trial function. The form of the first line in (\ref{lam20}) is chosen more or less arbitrarily. We demand only a smooth matching to the second line and exponential decay of $\lambda$ at $t\to-\infty$. The coefficient $k$ in the exponent $e^{kt}$ is chosen so that it is equal to the first decaying mode of the linearized Eq.~(\ref{H064}) at $t\to-\infty$.

We see that our OFM problem, defined by the equations~(\ref{H062}), (\ref{H064}), (\ref{H030}), (\ref{H040}), (\ref{H080}) and~(\ref{conservq}), is completely symmetric against the mirror mapping $x\leftrightarrow -x$. Hence, it is quite natural to investigate only symmetric solutions. Only such kind of solutions will be considered below. We may note additionally that the equation~(\ref{H064}) for $p$, considered in the backward direction in time $t$, with the `initial' condition~(\ref{H064}) does not depend at all on $q$ at given $L(t)$. We believe that the latter problem has only a symmetric solution, obeying~(\ref{H040}). We may recall that requirements of obeying Eq.~(\ref{H040}) demands a specific choice for $\Lambda$.

\subsubsection{Analytic self similar solutions}
\label{aps}

We consider in this subsection an approximate analytic solution of the problem with $\lambda(t)$ defined in Eq.~(\ref{lam20}) under condition that
\begin{equation}\label{H120}
t_\lambda\ll(-t)\ll1\,.
\end{equation}
In this case we may write instead of Eq.~(\ref{lam20})
\begin{equation}\label{H126}
\lambda(t)=-\frac{1}{4}\, \frac{\sqrt{t_\lambda}}{(t_\lambda-t)^{3/2}}\, ;
\end{equation}
and we may hope to find an analytic solution of our problem, at least at times (\ref{H120}). In the regime (\ref{H120}), we could neglect $t_\lambda$ in the denominator Eq.~\eqref{H126} and below in comparison to $(-t)$. However we leave it in this and analogous positions for clarity.

We will see that the solutions are composed of two parts. At $L(t)-x\gg\sqrt{-t}$, the solution is simple and very smooth. Such interval of the $x$-space we denote as $\bar{\Omega}$. At  $0<L(t)-x\lesssim\sqrt{-t}$, there is a somewhat nontrivial boundary layer. We will construct this part of the solution at the interval $\Omega$, corresponding to the condition: $0<L(t)-x\ll1$. We approximate the solution at $\Omega$ by a self similar solution, which will be described below. It is important that the domains $\bar{\Omega}$ and $\Omega$ are overlapping with each other at the interval $\sqrt{-t}\lesssim L(t)-x\ll 1$.

We will see below that 
\begin{equation}\label{H130}
-t \dot{L}(t) \ll \sqrt{-t}\, ,
\end{equation}
for the solution defined by Eq.~(\ref{H126}) under the condition~(\ref{H120}). We assume this strong inequality for now, and justify it a posteriori. The inequality~(\ref{H130}) means in particular that
the edge displacement $\bar{\ell}_0-L(t)$ is much less than the width of the boundary layer, where $\bar{\ell}_0-x\sim \sqrt{-t}$.

\begin{figure}[ht]
  \centering
  \includegraphics[width=8cm]{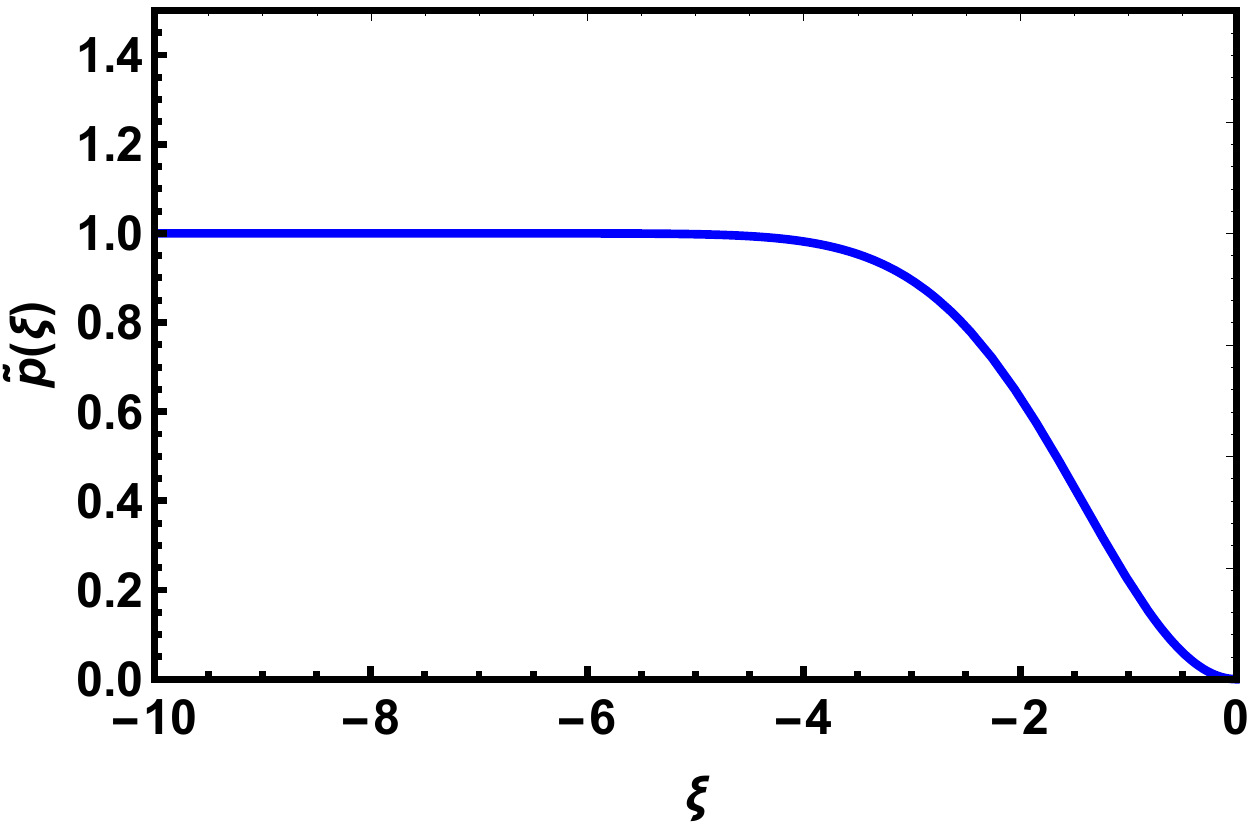}
  \caption{Shown is $\tilde{p}(\xi)$ defined in Eq.~(\ref{SS042}) versus $\xi$.}
  \label{figP}
\end{figure}

Eq.~(\ref{H064}) can be rewritten in the domain $\bar{\Omega}$ as
\begin{equation}\label{SS020}
\partial_t p=-\lambda(t)\, .
\end{equation}
Hence we have the following  solution for $p(x,t)$ in this domain:
\begin{equation}\label{SS030}
p(x,t)=p(0,t)=-\int\limits_{-\infty}^t\lambda(t)\, dt=\frac{1}{2}\,\sqrt{\frac{t_\lambda}{t_\lambda-t}}\, .
\end{equation}
Thus,
\begin{equation}\label{SS032}
\Lambda\simeq\frac{1}{2}\, .
\end{equation}
Taking in mind the strong inequality~(\ref{H130}), Eq.~(\ref{H064}) together with the boundary condition at  the swarm edge can be rewritten in the domain $\Omega$ as
\begin{equation}\label{SS034}
\partial_{t}p=-\partial_{x}^{2}p-\lambda(t)\,,\quad p\left(L(t),t\right)=0\,.
\end{equation}
Hence
\begin{equation}\label{SS040}
p(x,t)\simeq\frac{1}{2}\,\sqrt{\frac{t_\lambda}{t_\lambda-t}}
\left[1-\exp\left(-\frac{(x-L(t))^2}{4(t_\lambda-t)}\right)\right]
\end{equation}
for $x\in\Omega$.
The function
\begin{equation}\label{SS042}
\!\! \tilde{p}(\xi)=2\sqrt{\frac{t_{\lambda}-t}{t_{\lambda}}}p\left(\xi\sqrt{t_{\lambda}-t}+L(t),t\right)=1-e^{-\xi^{2}/4}
\end{equation}
is shown in Fig.~\ref{figP}. We see that the approximate solution inside the domain $\Omega$ has a self similar form. The function $\tilde{p}(\xi)$ represents this self similarity.

\begin{figure}[ht]
  \centering
  \includegraphics[width=8cm]{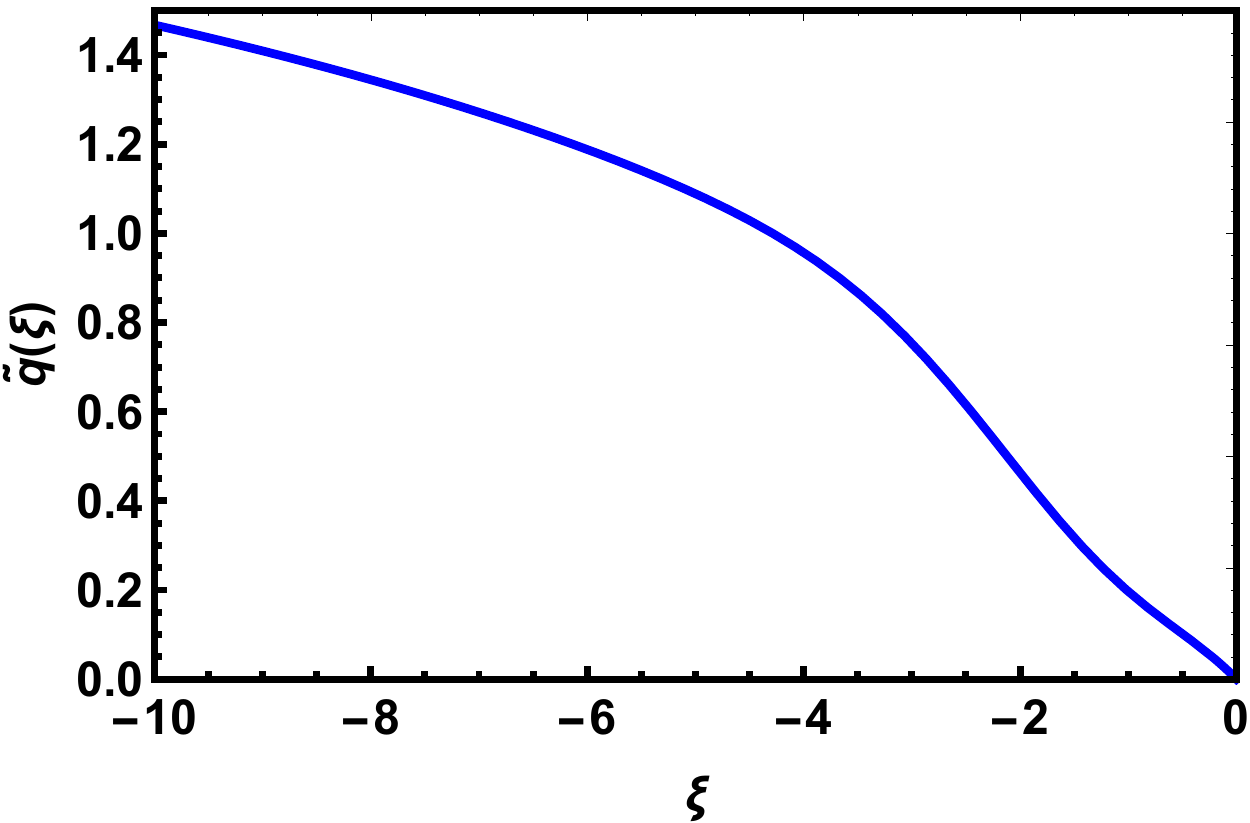}
  \caption{Shown is $\tilde{q}(\xi)$ defined in Eq.~(\ref{SS140}) versus $\xi$.}
  \label{figQ}
\end{figure}

Introducing $\tilde{q}_0(x,t)$ in the domain $\bar{\Omega}$ by the following definition
\begin{equation}\label{SS050}
q(x,t)=U(x)+\tilde{q}_0(x,t)\,,
\end{equation}
we obtain the following equation for $\tilde{q}_0(x,t)$ inside the domain $\bar{\Omega}$ from Eq.~(\ref{H062})
\begin{equation}\label{SS060}
\partial_t\tilde{q}_{0}=pU(x)=\frac{1}{2}\,\sqrt{\frac{t_\lambda}{t_\lambda-t}}U(x)\, .
\end{equation}
This equation may give only that $\tilde{q}_0(x,t)\sim \sqrt{t_\lambda} U(x)$ and it is determined by the time $(-t)\sim1$ that is outside the accuracy of the approximate solution considered in this section. In any case $\tilde{q}_0(x,t)/U(x)\sim \sqrt{t_\lambda}\ll 1$ in the domain $\bar{\Omega}$ at $t_\lambda\ll 1$. Inside the domain $\Omega$ we may use the following ansatz for $q$:
\begin{equation}\label{SS070}
q(x,t)\simeq\frac{L(t)-x}{2}+\tilde{q}\left(\frac{x-L(t)}{\sqrt{t_\lambda-t}}\right)\sqrt{t_\lambda}\,,
\end{equation}
where
\begin{equation}\label{SS074}
\dot{L}(t)=-\frac{\sqrt{t_\lambda}}{t_\lambda-t}\, f
\quad
(f=\mbox{const}\sim1)\,.
\end{equation}
The boundary condition~(\ref{H030}) and the conservation law~(\ref{conservq}) give the following boundary conditions for $\tilde{q}$:
\begin{eqnarray}
\label{SS080}
&& \tilde{q}(0)=0\,,\\
&& \!\!\!\! -2\left(\frac{\partial \tilde{q}}{\partial x}\right)_{x=L(t)}=-2\sqrt{\frac{t_\lambda}{t_\lambda-t}}\tilde{q}^\prime(0) \nonumber\\
\label{SS088}
&&\!\!\!\! =\int\limits_{-L(t)}^{L(t)} q(x,t) p(x,t)\, dx\simeq p(0,t)=\frac{1}{2}\,\sqrt{\frac{t_\lambda}{t_\lambda-t}}\,.
\end{eqnarray}
Hence,
\begin{equation}\label{SS090}
\tilde{q}^\prime(0)=-\frac{1}{4}\,.
\end{equation}
Substituting the ansatz~(\ref{SS070})-(\ref{SS074}) into Eq.~(\ref{H062}), we obtain the following  approximate equation for $\tilde{q}$:
\begin{equation}\label{SS100}
\frac{\dot{L}(t)}{2}+\partial_t\tilde{q}=\partial_x^2\tilde{q}-\partial_x \left((L(t)-x)\,\partial_x p\right)\,.
\end{equation}
This equation should be considered as linear relative to all perturbations of the equilibrium state. Using the expression~(\ref{SS040}) for $p$ inside the domain $\Omega$, we obtain that
the function $\tilde{q}(\xi)$ obeys the following ODE:
\begin{equation}\label{SS110}
-\frac{f}{2}+\frac{\xi}{2}\tilde{q}^\prime=\tilde{q}^{\prime\prime}+\frac{1}{4}\left(\xi^2e^{-\xi^2/4}\right)^\prime\,.
\end{equation}
Its unique solution obeying the condition~(\ref{SS080}), as well as a reasonable condition at $\xi\to-\infty$ can be presented as:
\begin{eqnarray}
\label{SS120}
\tilde{q}(\xi)&=&-\frac{f}{4}\left[\xi^2\, _2F_2\left(\{1,1\},\left\{\frac{3}{2},2\right\},\frac{\xi^2}{4}\right)+2\pi\mbox{erfi}\left(\frac{\xi}{2}\right)\right] \nonumber\\
&+&\frac{\xi}{4}e^{-\xi^2/4}\,.
\end{eqnarray}
Here $_mF_n(\dots)$ and $\mbox{erfi}(.)$ are the generalized hypergeometric function and the imaginary error function, respectively~\cite{dlmf}. We have
\begin{equation}\label{SS130}
\tilde{q}(\xi\to 0)=-\frac{f\sqrt{\pi}}{2}\xi+\frac{1}{4}\xi\,.
\end{equation}
This equation together with the boundary condition~(\ref{SS090}), coming from conservation of total number of bees, give
$$
f=\frac{1}{\sqrt{\pi}}\, .
$$
Plugging this into Eq.~\eqref{SS120}, we obtain
\begin{eqnarray}
\label{SS140}
\tilde{q}(\xi) \!\! &=& \!\!-\frac{1}{4\sqrt{\pi}} \! \left[\xi^2\, _2F_2\!\left(\{1,1\},\left\{\frac{3}{2},2\right\},\frac{\xi^2}{4}\right) \! +2\pi\mbox{erfi} \! \left(\!\frac{\xi}{2}\!\right)\right] \nonumber\\
&+&\!\! \frac{\xi}{4}e^{-\xi^2/4}\,,
\end{eqnarray}
and
\begin{equation}\label{SS160}
\dot{L}(t)=-\frac{\sqrt{t_\lambda/\pi}}{t_\lambda-t}\,.
\end{equation}
The function $\tilde{q}(\xi)$ is shown in Fig.~\ref{figQ}. At $\xi\to-\infty$, it behaves as:
\begin{equation}\label{SS164}
\tilde{q}(\xi)\to \frac{1}{2\sqrt{\pi}}\, \left(\ln \xi^2+\gamma_E\right),
\quad (\xi\to-\infty)
\end{equation}
where $\gamma_{E} = 0.577\dots$ is the Euler constant.

Thus, Eqs.~(\ref{SS030}), (\ref{SS040}), (\ref{SS050})-(\ref{SS070}), (\ref{SS140}) and~(\ref{SS160}) give complete description of our solution at $t_\lambda\to 0$ and under the condition~(\ref{H120}). It obeys also the condition~(\ref{H130}) that was used implicitly several times during this derivation.

\subsubsection{Calculation of the action for the analytic solution}
\label{AA}

Eq.~(\ref{H090}) gives the following expression for the action rate $\dot{S}$ at $t_\lambda\ll(-t)\ll1$
\begin{equation}\label{SS170}
\dot{S}=\frac{1}{2}\int\limits _{-L(t)}^{L(t)}qp^{2}\,dx+\int\limits _{-L(t)}^{L(t)}q\left(\partial_{x}p\right)^{2}\,dx\,,
\end{equation}
because of the condition $p\ll 1$ during this period.
The main contribution to the 1st integral comes from the domain $\bar{\Omega}$. Thus, it can be calculated as
\begin{equation}\label{SS180}
\!\dot{S}_1=\frac{1}{2} \! \int\limits_{\bar{\Omega}} \! qp^2\, dx \simeq \frac{p^2(0,t)}{2}\int\limits_{-\bar{\ell}_0}^{\bar{\ell}_0} \! U(x)\, dx=\frac{1}{8}\, \frac{t_\lambda}{t_\lambda-t}\,.
\end{equation}
The main contribution to the 2nd integral comes from the the domain $\Omega$:
\begin{eqnarray}
\label{SS190}
\dot{S}_2&=&\int\limits_{\Omega} qp_x^2\, dx \nonumber\\
&\simeq& 2 \int\limits_{0}^{L(t)} \frac{L(t)-x}{2} \frac{t_\lambda}{4(t_\lambda-t)}\frac{(x-L(t))^2}{4(t_\lambda-t)^2} \nonumber\\
&\times& \exp\left(-\frac{(x-L(t))^2}{2(t_\lambda-t)}\right) \, dx \, .
\end{eqnarray}
Making here the substitution $(L(t)-x)/\sqrt{t_\lambda-t}=u$, we obtain:
\begin{equation}\label{SS200}
\dot{S}_2\simeq\frac{t_\lambda}{16(t_\lambda-t)} \int_0^\infty u^3\, e^{-u^2/2}\, du =\frac{t_\lambda}{8(t_\lambda-t)} \,.
\end{equation}
Combining both contributions to the action rate we obtain:
\begin{equation}\label{SS210}
\dot{S}=\dot{S}_1+\dot{S}_2=\frac{t_\lambda}{4(t_\lambda-t)} \,.
\end{equation}

Thus we have expressions for $\dot{L}$ and $\dot{S}$ for our solution at $1\ll -t\ll t_\lambda$. See Eqs.~(\ref{SS160}) and~(\ref{SS210}). We see that the total edge displacement $\Delta L$ as well the action $S$ diverge at $t_\lambda-t\to 0$ and $\to\infty$ if we extend the expressions~(\ref{SS160}) and~(\ref{SS210}) outside their domain of applicability,  $1\ll -t\ll t_\lambda$. It is a key point of the OFM theory for this system at $\Delta L\to0^-$.   This property allows us to make integration over time interval $t\in(-1+t_\lambda,0)$ to get approximate evaluation of the whole action $S$ and the whole edge displacement $\Delta L$. The times $-t\gtrsim 1$ and $-t\lesssim t_\lambda$ give some contributions to these values that can be estimated  as $\sim\sqrt{t_\lambda}$ and $\sim t_\lambda$, respectively. They can be neglected in the leading order for $\Delta L$ and $S$ due to the `divergences' mentioned above. This assumption will be confirmed in Sec.~\ref{nps} by a direct simulation of our  whole problem.  As a result, we have at $t_\lambda\to 0$:
\begin{equation}\label{SS212}
\Delta L = \sqrt{\frac{t_\lambda}{\pi}}\, \bigl[\ln t_\lambda +{\cal O}(1)\bigr]\, ,
\end{equation}
and
\begin{equation}\label{SS214}
S = \frac{t_\lambda}{4}\left[\ln\frac{1}{t_\lambda} +{\cal O}(1)\right]\, ,
\end{equation}
These two equations give the following relationship:
\begin{equation}\label{SS230}
S\frac{|\ln \Delta L^2|}{\Delta L^2}=\frac{\pi}{4}\left[1-2\frac{\ln\ln\frac{1}{t_\lambda}}{\ln\frac{1}{t_\lambda}}+{\cal O}\left(\frac{1}{\ln\frac{1}{t_\lambda}}\right)\right]\, .
\end{equation}
For $t_\lambda\to 0$ (and hence, $\Delta L\to0^-$) we obtain from the latter equation:
\begin{equation}\label{SS220}
S = \frac{\pi}{4} \frac{\Delta L^2}{|\ln \Delta L^2|}+\dots
\qquad (\Delta L\to 0^-) \,.
\end{equation}
This is our main result for $\lambda(t)$ given by Eq.~(\ref{H126}) at times $1\gg -t\gg t_\lambda$.

\subsubsection{Numerical solution for the chosen $\lambda(t)$}
\label{nps}

We solve numerically the problem~(\ref{H062}), (\ref{H064}), (\ref{H030}), (\ref{H040}), (\ref{H080}) and~(\ref{conservq}) at known $\lambda(t)$, given by Eq.~(\ref{lam20}). We replace the boundary condition~(\ref{H040}) at infinite past on the boundary condition at finite time $t=-T$
\begin{equation}
\label{SS280}
q(x,-T)=U(x)\, ;\quad
p(x,-T)=0\, ,
\end{equation}
 where $T > 0$ is sufficiently large to exclude influence of finiteness of $T$ on our solutions. Rigorously speaking, we cannot demand $p(x,-T)=0$ for finite $T$. Actually, we replace the second condition in~(\ref{SS280}) by demanding that
\begin{equation}
\label{SS282}
p(0,-T)=0 \, .
\end{equation}
This condition can be fulfilled at a specific value of $\Lambda$ as can be seen from the set of equations~(\ref{H064}), (\ref{H080}) and~(\ref{SS280}). The latter value is actually an eigenvalue of the problem, that depends on $\lambda(t)$. To be sure that our choice of $T$ is large enough to approximate well the solution of the original problem (with $T\to\infty$), we calculate for this specific value of $\Lambda$ the integral $\int_{-\bar{\ell}_0}^{\bar{\ell}_0} p^2(x,-T)\, dx$. We checked that this integral would be sufficiently small for our choice of $T$.

We solve this boundary value problem  with the iteration procedure, ascending to the work~\cite{ChSt}. The parameter $\Lambda$ plays a role of an eigenvalue. Each step of the iteration consists of two sub-steps: i) forward and ii) backward.  The equation~(\ref{H062}) for $q$ is solved forward in time with the boundary condition~(\ref{SS280}) treated as an initial condition. We use at this substep the function $p$, obtained during the previous iteration step. Then the equation~(\ref{H064}) is solved backward in time for $p$ using the boundary condition~(\ref{H080}) as an initial condition. Our present problem has a novel feature. It is a free boundary problem with unknown in advance $L(t)$. We calculate $L(t)$ in the forward sub-step of the iteration procedure using the condition~(\ref{conservq}) simultaneously with solving of Eq.~(\ref{H062}). Details of this procedure are given in Appendix \ref{AppendixNumerical}. During the backward sub-step of the iteration we used $L(t)$ obtained at the previous forward sub-step.

This method allows us to find the solution with given $\lambda(t)$ and $\Lambda$. It does not satisfy yet the condition~(\ref{SS282}) for $p$. Using the method described in the previous paragraph, we apply a shooting procedure to find such value of $\Lambda$, which corresponds to a solution obeying Eq.~(\ref{SS282}).

\begin{figure*}[ht]
  \centering
  \includegraphics[width=8cm]{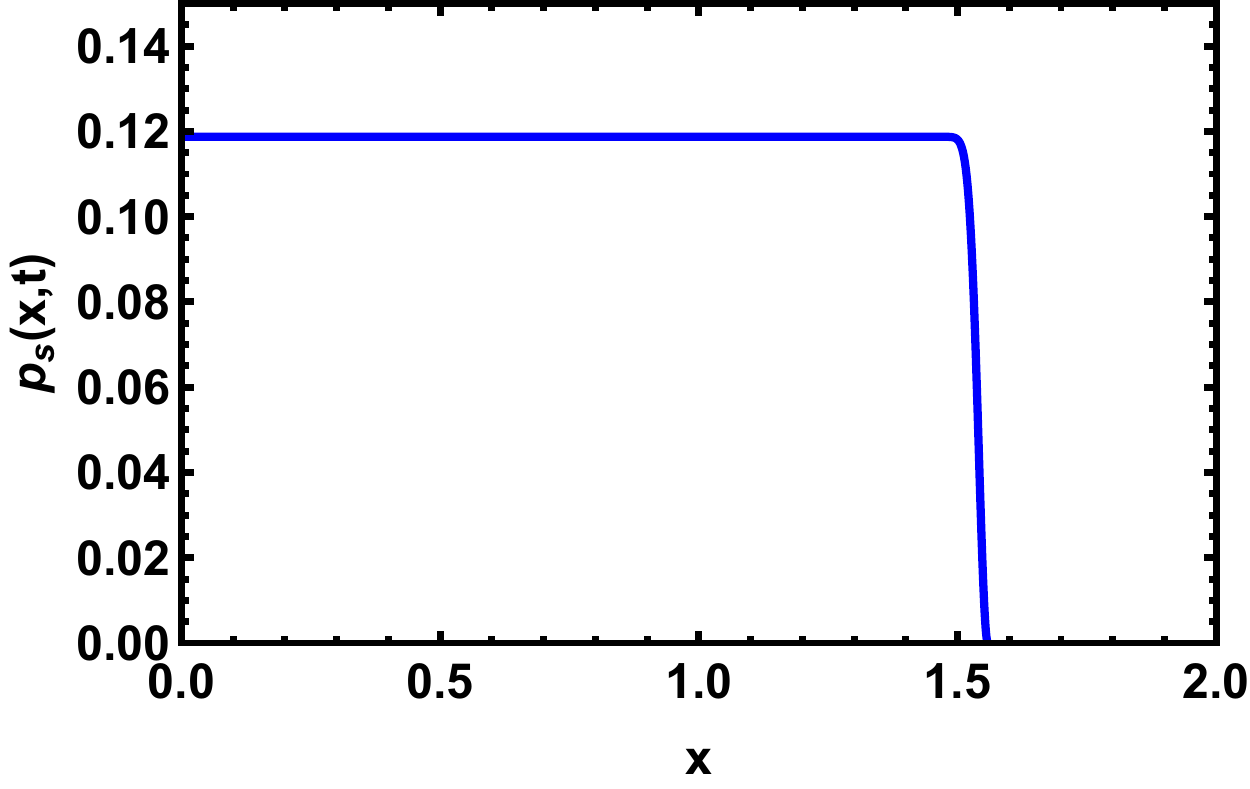}
  \includegraphics[width=8cm]{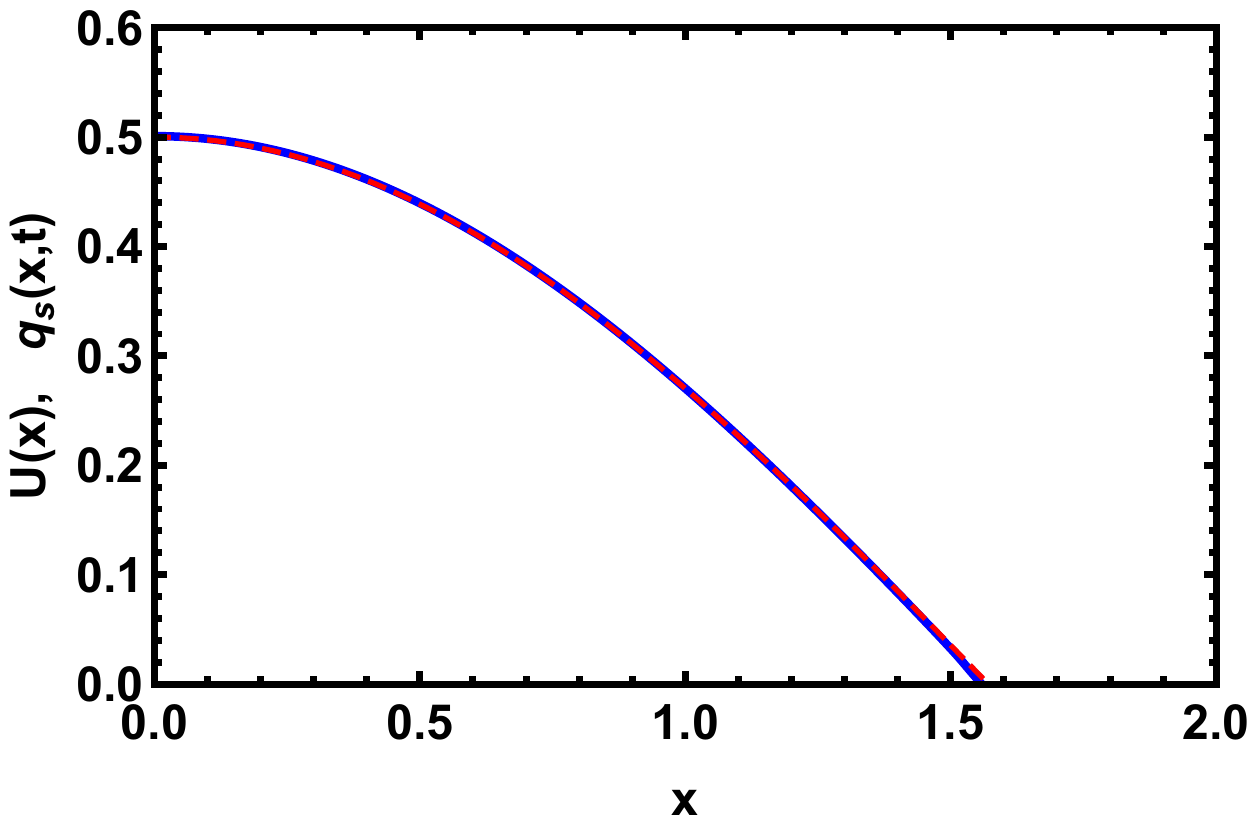}
  \caption{Numerical solution of the OFM problem with $t_\lambda=10^{-5}$ at $t=-15.7\, t_\lambda$. The solution in the bulk, in the domain $\bar{\Omega}$ is shown here. The left panel shows $p_s(x,t)$. Compare it with Eq.~(\ref{SS030}) that gives $p(x\in\bar{\Omega},t)\simeq 0.122$ at this time. The right panel shows $q_s(x,t)$ by the blue line and $U(x)$ by the red dashed line. See Eq.~(\ref{SS050}). The difference between the latter two lines is almost invisible, excluding the region near $x=\bar{\ell}_0$ (the domain $\Omega$).}
  \label{bulk}
\end{figure*}

Important details providing the solution of the problem~(\ref{H062}), (\ref{H064}), (\ref{H030}), (\ref{H040}), (\ref{H080}) and~(\ref{conservq}) at known $\lambda(t)$ are given in Appendix \ref{AppendixNumerical}. We applied this method for the one-parametric set~(\ref{lam20}) of functions $\lambda(t)$ with $t_\lambda =0.1$, $0.02$, $10^{-2}$, $10^{-3}$, $10^{-4}$, $10^{-5}$ and $10^{-6}$.

\begin{figure*}[ht]
  \centering
  \includegraphics[width=8cm]{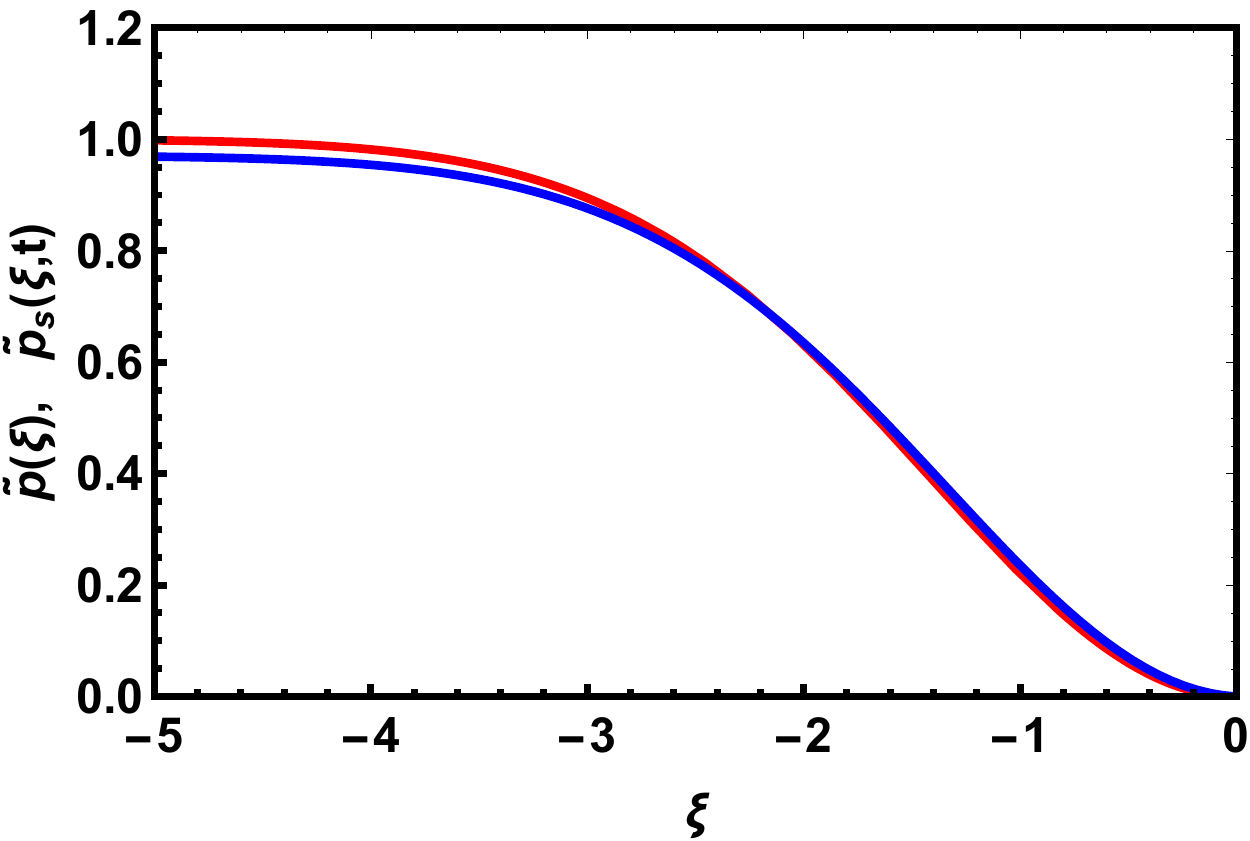}
  \includegraphics[width=8cm]{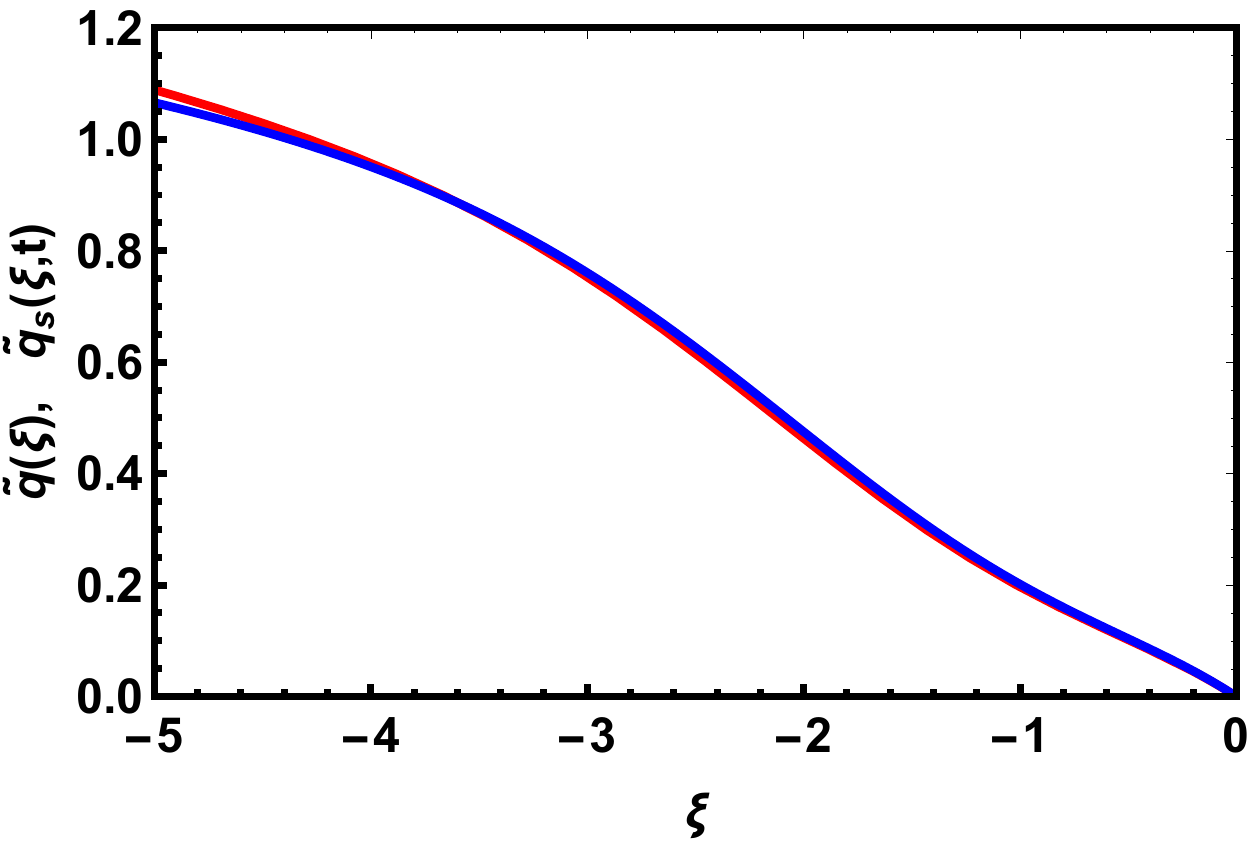}
  \caption{Shown is a comparison of the (theoretical) self-similar and numerical solutions. The left panel shows the momentum in the domain $\Omega$.  The right panel shows the normalized bees' density $q$.  The blue curves are obtained from numerical results for $t_\lambda=10^{-5}$ and $t=-15.7\, t_\lambda$ in accordance to Eqs.~(\ref{SS310}) and~(\ref{SS300}), respectively. The red lines are the self-similar solutions~(\ref{SS040}) and~(\ref{SS140}), respectively.}
  \label{compP}
\end{figure*}

First of all, we compare the simulated $q_s(x,t)$ and $p_s(x,t)$ with the self similar solution considered in Sec.~\ref{aps} at small $t_\lambda$. We designate the simulated $q$ and $p$ by the subscript `$s$'. Figs.~\ref{bulk} and~\ref{compP} show the numerical solution with $t_\lambda=10^{-5}$ at $t=-15.7\, t_\lambda = 1.57\times10^{-4}$. Fig.~\ref{bulk} shows the numerical solution in the bulk, in the domain $\bar{\Omega}$.

To compare the self-similar like solution (in the domain $\Omega$) to the simulations we may use the following expressions 
\begin{equation}\label{SS300}
\tilde{q}_s(\xi,t)=\frac{1}{\sqrt{t_\lambda}}q_s\left(L(t)+\xi\sqrt{t_\lambda-t},t\right)+\xi\sqrt{\frac{t_\lambda-t}{4t_\lambda}}\,, 
\end{equation}
and
\begin{equation}\label{SS310}
\tilde{p}_s(\xi,t)=\sqrt{\frac{4(t_\lambda-t)}{t_\lambda}}p_s\left(L(t)+\xi\sqrt{t_\lambda-t},t\right)\,.
\end{equation}
Compare these expressions with  Eq.~(\ref{SS070}) and Eqs.~(\ref{SS040}) and~(\ref{SS042}), respectively. These expressions should coincide with the self similar analytic solutions $\tilde{q}(\xi)$ and $\tilde{p}(\xi)$ of Sec.~\ref{aps}, respectively, at $1\ll (-t)\ll t_\lambda$ and $t_\lambda\to0$.

Comparison of $\tilde{p}_s(\xi,t)$ with theoretical $\tilde{p}(\xi)$ is presented in Fig.~\ref{compP}(left panel). The curve $\tilde{p}_s(\xi,t)$, presented there, is obtained from the simulation with  $t_\lambda=10^{-5}$ and $t=-15.7\, t_\lambda$. We use the expression~(\ref{SS310}) to calculate $\tilde{p}_s(\xi,t)$. Analogously, Fig.~\ref{compP}(right panel) shows comparison of $\tilde{q}_s(\xi,t)$ with the theoretical curve $\tilde{q}(\xi)$. The same simulated data, but for $q_s(x,t)$ is used to calculate $\tilde{q}_s(\xi,t)$ in accordance to Eq.~(\ref{SS300}).  We see that the correspondence  between simulated and self-similar solutions is quite acceptable. We checked that the same statement is valid  for all times from the interval $0.03\lesssim(-t)\lesssim t_\lambda$ for sufficiently small $t_\lambda$ (not shown). This means that indeed the self similar solutions, considered in Sec.~\ref{aps}, correctly describe the intermediate asymptotic behavior of the full OFM solutions corresponding to the set of $\lambda(t)$, defined in Eq.~(\ref{lam20}), at $t_\lambda\to0$.

\begin{figure}[ht]
  \centering
\includegraphics[width=8cm]{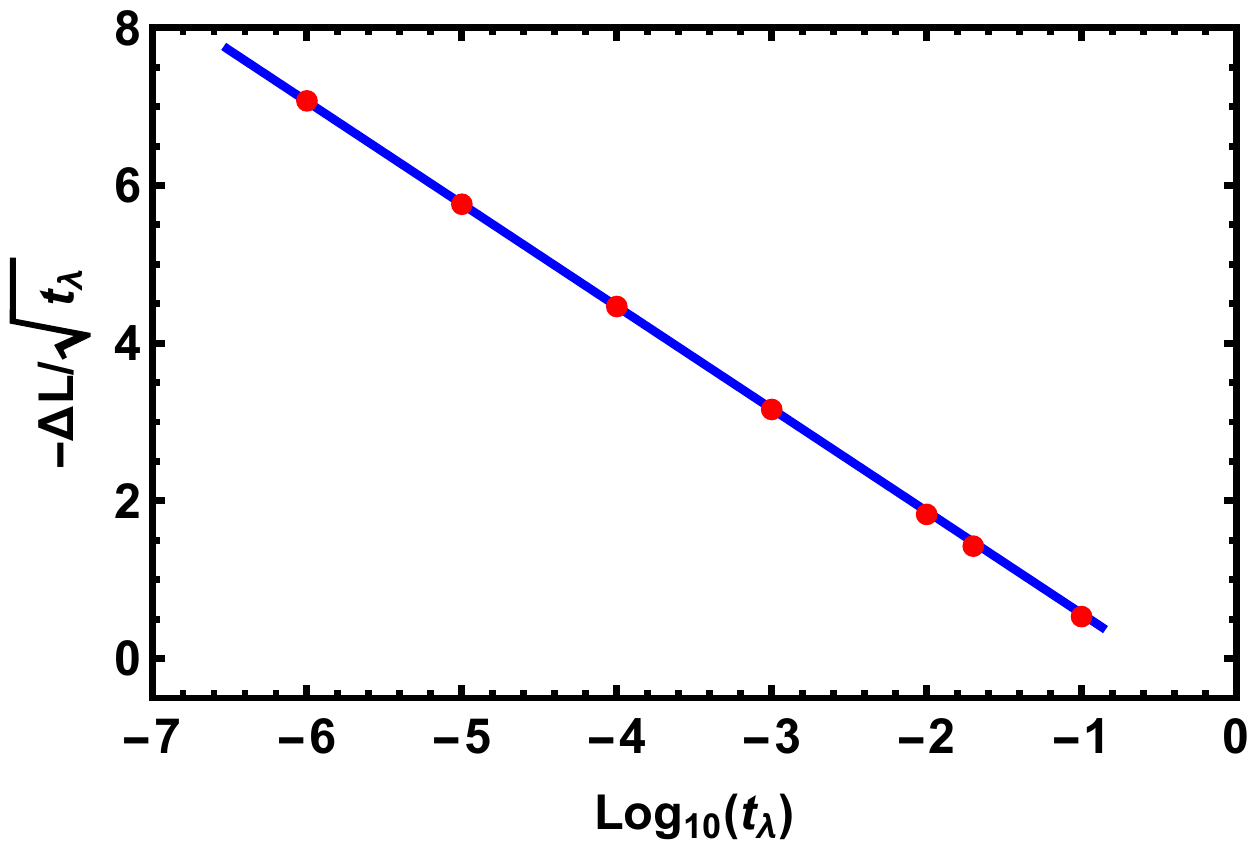}
  \caption{Simulated relationship between the edge displacement  $\Delta L$ and $t_\lambda$. The red points represent results of the simulations.  The blue line shows the fit~(\ref{SS380}). See also Eq.~(\ref{SS212}).}
  \label{dlfig}
\end{figure}
\begin{figure}[ht]
  \centering
  \includegraphics[width=8cm]{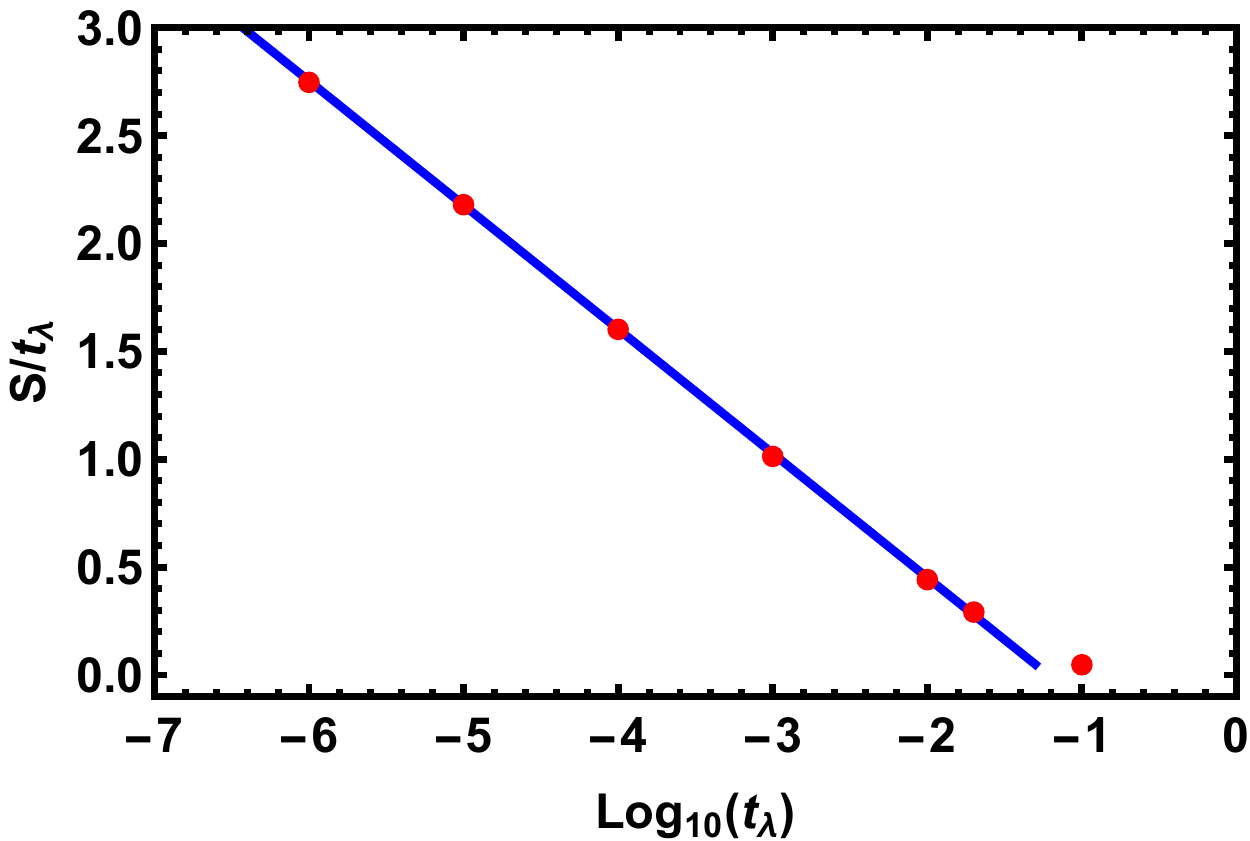}
  \caption{Simulated relationship between the action $S$ and $t_\lambda$. The red points represent results of the simulations.  The blue line shows the fit~(\ref{SS390}). See also Eq.~(\ref{SS214}).}
  \label{sfig}
\end{figure}

Now we proceed to analysis of the integral parameters of the numerical solutions, total action, $S$, and total edge displacement, $\Delta L$. They depend now on $t_\lambda$ only. Figs.~\ref{dlfig} and~\ref{sfig} show comparison of dependencies of normalized total displacements $\Delta L/\sqrt{t_\lambda}$ and normalized total actions $S/t_\lambda$ on $t_\lambda$ with fits based on Eqs.~(\ref{SS212}) and~(\ref{SS214}).   To get self-similar theoretical results~(\ref{SS212}) and~(\ref{SS214}) we integrated the expressions~(\ref{SS160}) and~(\ref{SS210}) over $t$ formally from $t=-1+t_\lambda$ to $t=0$. Keeping in mind an analytically uncertain contributions to these integrals from the regions $(-t)\gtrsim 1$ and $0<(-t)\lesssim t_\lambda$, we may suppose the existence of analytically uncertain  constant multipliers of the order of 1 under the logarithms. We add such multipliers to fit the simulated data. The blue lines in Figs.~\ref{dlfig} and~\ref{sfig} correspond to a specific choice of these factors.  Actually the blue line in Fig~\ref{dlfig} corresponds to the  relationship
\begin{equation}\label{SS380}
\Delta L=\sqrt{\frac{t_\lambda}{\pi}}\, \ln 3.65 t_\lambda\, .
\end{equation}
Comparing it with Eq.~(\ref{SS212}), we may conclude that the numerical solutions confirm existence of the logarithm and even the coefficient  $1/\sqrt{\pi}$ in Eq.~(\ref{SS160}), when $t_\lambda\ll 1$. At $t_\lambda\sim 10^{-6} - 10^{-5}$ the contribution of the multiplier 3.65 to the leading order becomes really small. Analogously, the blue line in Fig.~\ref{sfig} corresponds to the  relationship
\begin{equation}\label{SS390}
S=\frac{t_\lambda}{4}\, \ln \frac{0.061}{t_\lambda}\, .
\end{equation}
It can be compared with Eq.~(\ref{SS214}). Although, the subleading term determined by the factor 0.061 is relatively higher than the analogous correction in Eq.~(\ref{SS390}), nevertheless the leading order term dominates at $t_\lambda\sim 10^{-6} - 10^{-5}$. Thus, the numerical solutions confirm the existence of the logarithm multiplier in the asymptotics~(\ref{SS214}) as well as the overall coefficient $1/4$ in it.

\begin{figure}[ht]
  \centering
  \includegraphics[width=8cm]{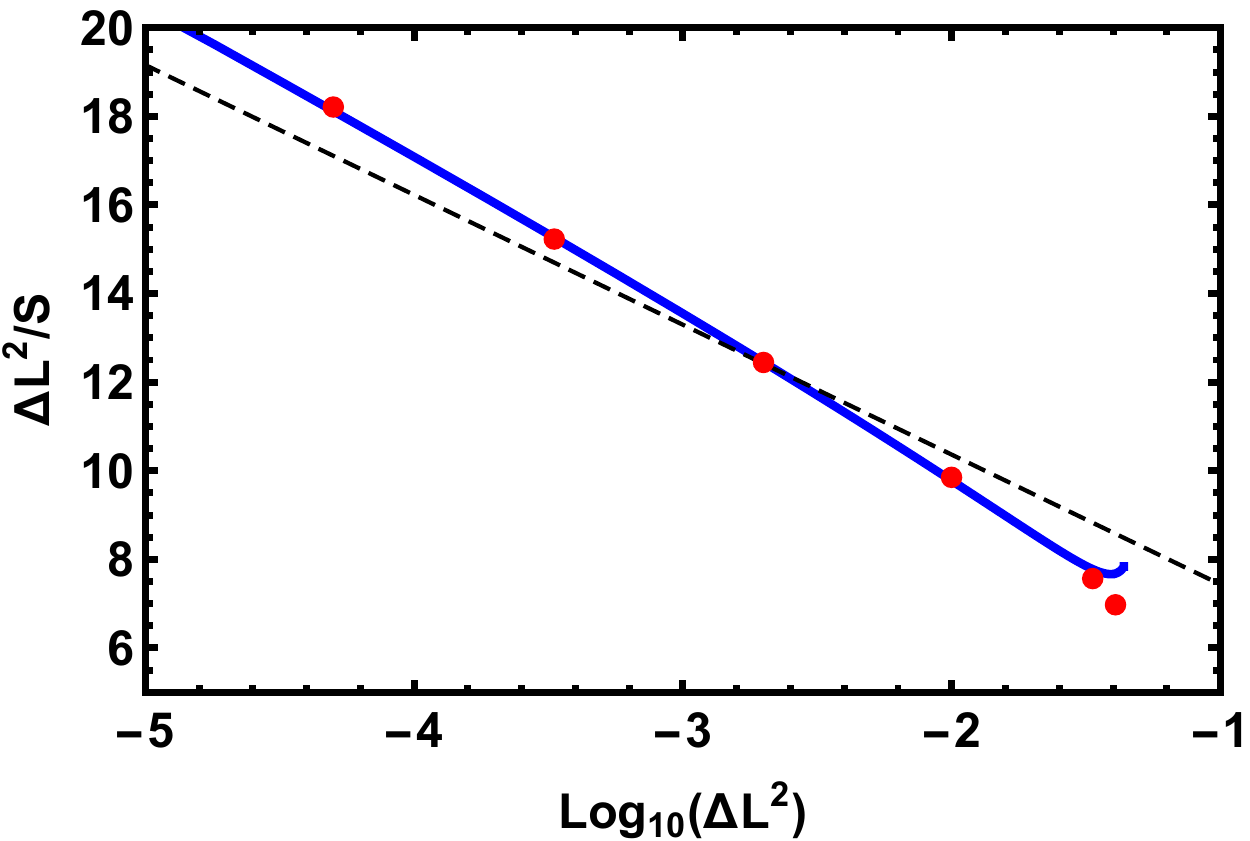}
  \caption{The blue line is a curve defined parametrically by Eqs.~(\ref{SS380}) and~(\ref{SS390}). The red points are obtained from  results of the numerical solutions. The dashed black line corresponds to $S=(\pi/4)\Delta L^2/|\ln \left(0.011\Delta L^2\right)|$. Existence of subleading terms, considered in the text, are seen apparently. }
  \label{SL}
\end{figure}

The dependence $S(\Delta L)$ defined parametrically by Eqs.~(\ref{SS380}) and~(\ref{SS390}) is plotted in Fig.~\ref{SL}. It shows also points obtained from results of the simulations. We see that subleading terms, caused by the approximate exclusion of $t_\lambda$ from Eqs.~(\ref{SS380})-(\ref{SS390}) [or from Eqs.~(\ref{SS212})-(\ref{SS214})],  are well seen at our $\Delta L$ (or $t_\lambda$). To reveal this fact analytically we may consider Eq.~(\ref{SS230}). 
The 3rd term in the right hand side of this equation is of the same order of what would give subleading order terms in Eqs.~(\ref{SS212}) and~(\ref{SS214}). We see their relative contributions decay only logarithmically at $t_\lambda\to 0$. Nevertheless, the 2nd term in the right hand side of Eq.~~(\ref{SS230}) is decaying even slower.
We see that to reach the region where the subleading term would be about 10\% of the leading term in this equation we should set $t_\lambda\lesssim 10^{-19}$. Such values seems to us as unreachable for our present numerical methods. Nevertheless, we may say that the simulations confirm surely the asymptotic behaviors~(\ref{SS212}) and~(\ref{SS214}) at $t_\lambda\to0$ for the set~(\ref{lam20}) of the functions $\lambda(t)$.

Combining together all results of Secs.~\ref{aps}-\ref{nps}, we may draw the following conclusions about the OFM solutions with $\lambda(t)$, defined by Eq.~(\ref{lam20}) and parameterized by $t_\lambda$, tending to 0 : 
\begin{itemize}
\item The main contributions to the displacement of the edge and to the action come from the time interval $t_\lambda\ll (-t)\ll 1$.
\item The OFM solutions at $t_\lambda\ll (-t)\ll 1$ can be well described by the self similar solutions investigated in Secs.~\ref{aps} and~\ref{AA}.
\item As a result, the action, $S$, on this set of solutions can be described by Eq.~(\ref{SS220}) at $\Delta L\to 0^-$.
\item Non self similar contributions to the displacement, $\Delta L$, and to the action, $S$, influence only  subleading terms in Eq.~(\ref{SS220}).
\end{itemize}

\subsection{General remarks about the OFM solution}
\label{GR}

We explain in Sec.~\ref{ps} that the one parametric set of the functions $\lambda(t)$, determined by Eq.~(\ref{lam20}),  gives the asymptotic relationship between the action, $S$, and the edge displacement, $\Delta L$, presented by Eq.~(\ref{SS220}). Since this relationship corresponds to a particular choice of the set of the functions $\lambda(t)$, we may conclude that the Eq.~(\ref{SS220}) gives only an upper bound for $S(L)$ at $\Delta L\to 0^-$. Nevertheless, we present in this section arguments in favor of the claim that the specific behavior of the functions $\lambda(t)$ from this set at $t_\lambda\ll (-t)\ll 1$, when $t_\lambda$ tends to 0, provides the valid asymptotic  leading term in this expression as a solution of the OFM problem, described in Sec.~\ref{GE}. The particular form of $\lambda(t)$ outside the time interval $t_\lambda\ll (-t)\ll 1$ determines only subleading terms in Eq.~(\ref{SS220}), but not the leading term. The subleading terms are neglected by us in Eq.~(\ref{I040}). A key point for such conclusion is that $S(L)/\Delta L^2$ tends to 0 at $\Delta L\to 0^-$. This property could be valid only  for a quite specific choice of the set of the functions $\lambda(t)$; and our choice of $\lambda(t)$ in Sec.~\ref{ps} ensures such specific properties.

We start our way to the set~(\ref{lam20}) of the functions $\lambda(t)$ from several examples of trial functions that show us how to obtain
\begin{equation}\label{E004}
\frac{S(L)}{\Delta L^2}\to0\qquad\mbox{~at~~~~~} \Delta L\to 0^-\,.
\end{equation}

\subsubsection{The case of single time scale}
\label{1sc}

Let us consider firstly functions $\lambda(t)$ that have only a single time scale, $t_0$. We focus  on the behavior of $p(0,t)$ and do not specify exactly $\lambda(t)$ in Eq.~(\ref{H064}) providing such $p(0,t)$. For $0<(-t)<t_0$ we set some $p(0,t)\sim p_0>0$, and set that $p(0,t)$ [as well as $p(x,t)$] tends quickly to 0 for larger $(-t)$. We consider below several combinations of $t_0$ and $p_0$ detemined by strong inequalities.

\paragraph*{\underline{The case of $t_0\gg1$.}} -- In this case we may use results of Ref.~\cite{MS21}, devoted to persistent fluctuations in the Brownian bees model. We obtain that
$$
S(L)\sim \Delta L^2\, t_0
$$ 
for such trial functions. We may conclude that we should set $t_0$ as small as possible, while $t_0\gg 1$, and that even for smallest possible $t_0\sim 1$, we have $S\sim \Delta L^2$. The latter estimation is much higher than in Eq.~(\ref{SS220}) and does not obey the condition~(\ref{E004}). 


Considering the case of $t_0\ll1$, we separate it into two limiting subcases: i) $p_0\ll1$; and ii) $p_0\gg 1$.

\begin{figure}[ht]
  \centering
  \includegraphics[width=8cm]{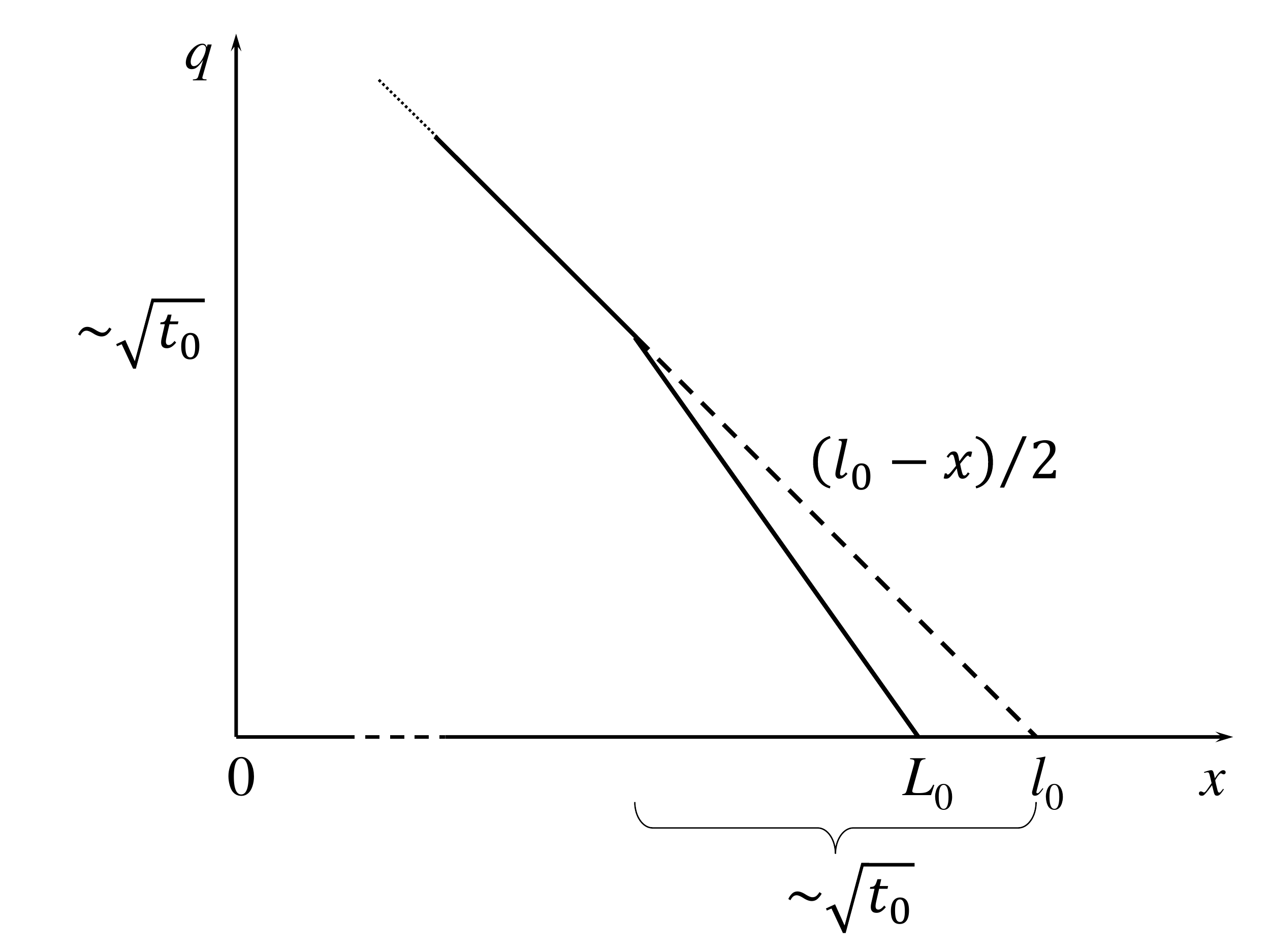}
  \caption{Shown is a sketch for $q(x,0)$ in the case of single time scale $\lambda(t)$, when $p(0,0)$ and the time scale $t_0 \ll 1$. It corresponds to the solid line. The region $\bar{\ell}_0-x \ll 1$ is shown only. The dashed line presents $U(x)$. The aria between the dashed and solid lines can be estimated as $\sim (\bar{\ell}_0-L_0)\sqrt{t_0}$.}
  \label{sk1}
\end{figure}

\paragraph*{\underline{The case of $t_0\ll1$ and $p_0\ll1$.}} -- We begin from the case (i), when $p_0\ll1$. We will see that in this case absolute value of the edge displacement $\Delta L = L_0-\bar{\ell}_0$ is much less than the width of the diffusion boundary layer, that can be estimated as $\sqrt{t_0}$. This boundary layer corresponds actually to the domain $\Omega$ introduced above. In this case the density distribution in $\Omega$ is similar to a quasi-equilibrium distribution, which is adjusted to a new position of the swarm edge. See Fig.~\ref{sk1}. The fields $p$ and $q$ outside this boundary layer [that corresponds actually to the domain $\bar{\Omega}$ introduced above] are not affected by the diffusion process during the time, when $(-t)\sim t_0$, besides a small increasing of $q$ in this domain due to positive $p$. As a result, $p(x,t)$ is constant in space in the domain $\bar{\Omega}$. The total normalized number of particles will be almost the same as in the equilibrium with a small excess $\Delta Q=\int_{\bar{\Omega}} (q-U)\, dx$, which can be estimated as $\Delta Q\sim p_0 t_0$. This excess in the bulk (in the domain $\bar{\Omega}$) should be compensated by a lack of particles in the domain $\Omega$. See for clarification Fig~\ref{sk1}. The lack $\Delta Q_e$ can be estimated as an area between the dashed and solid lines in Fig.~\ref{sk1}. The former one represents initial equilibrium, $U(x)$ close to the edge, whereas the latter one represents perturbed $q(x,0)$ close to the edge. This area can be estimated as $|\Delta L| \sqrt{t_0}$. Hence, $\Delta Q_e\sim|\Delta L| \sqrt{t_0}$. Owing to conservation of total number of particles, we should demand at least $\Delta Q\sim \Delta Q_e$, and, hence $\Delta L \sqrt{t_0}\sim - p_0 t_0$. As a result, we have the following estimation of $\Delta L$ in this case:
\begin{equation}\label{E010}
\Delta L \sim - p_0 \sqrt{t_0}\,.
\end{equation}
This estimations confirms our preliminary assumption that $|\Delta L|\ll \sqrt{t_0}$ in this case.

Let us estimate now the action for the case $p_0\ll1$ and $t_0\ll1$. Both contributions to the action in Eq.~(\ref{H090}) are determined by the time interval $(-t)\sim t_0$. The 1st contribution is determined by the domain $\bar{\Omega}$, whereas the 2nd one by the contribution from the diffusion boundary layer. It appears that the both contributions can be estimated as $p_0^2 t_0$. Hence,
\begin{equation}\label{E020}
S \sim  p_0^2 t_0\,.
\end{equation}

Combining Eqs.~(\ref{E010}) and~(\ref{E020}), we obtain the following estimation for the action $S$:
\begin{equation}\label{E030}
S \sim \Delta L^2\,.
\end{equation}
First of all, we note that this action at $\Delta L\to 0$ 
does not obey the condition~(\ref{E004}).  We may note also that the action do not depend on $t_0$ or $p_0$ but only on their combination appearing in Eq.~(\ref{E010}). This fact will be important below.

\paragraph*{\underline{The case of $t_0\ll1$ and $p_0\gg1$.}} -- This case differs from the previous one in two points. The additional to the equilibrium normalized number of particles begotten during the period $(-t)\sim t_0$ can be estimated now as $\Delta Q \sim  e^{p_0}t_0$, whereas additional lack of particles near the edge can be estimated now as $\Delta Q_e\sim\Delta L^2$. The latter estimation comes from a reasonable assumption that the perturbation of the slope of $q(x,0)$ in the boundary layer is of the order of the slope in the equilibrium. Thus, due to conservation law ($\Delta Q\sim \Delta Q_e$), we have:
\begin{equation}\label{E040}
\Delta L^2 \sim e^{p_0} t_0\,.
\end{equation}
Only the 1st term in Eq.~(\ref{H090}) gives considerable contribution to the action. As a result we have:
\begin{equation}\label{E050}
S \sim p_0 e^{p_0} t_0\,.
\end{equation}
Combining these two equations we obtain for the present case:
\begin{equation}\label{E060}
S \sim p_0 \Delta L^2\qquad (p_0\gg 1)\,.
\end{equation}
We see that the lowest possible action for $p_0\gtrsim 1$ takes place at $p_0\sim 1$. Again even in the latter case ($p_0\sim 1$) this action at $\Delta L \to 0$ is much higher than the action~(\ref{SS220}) for particular solutions, considered in Sec.~\ref{ps}.

We may draw the following general conclusion for the cases of a single time scale trial functions $\lambda(t)$. Such trial functions give that $S\sim\Delta L^2$ or higher. In any case the action becomes much higher than the action~(\ref{SS220}) at $\Delta L\to 0^-$ for particular solutions, considered in Sec.~\ref{ps}.

\subsubsection{Multi scale in time trial functions}

Before turning to power law form of trial functions $\lambda(t)$, which could be a candidate for the multi scale in time trial functions, we consider in more details a degeneracy revealed when we considered the case $t_0$ and $p_0\ll1$ in Sec.~\ref{1sc}. We saw there that any time interval $(-t)\sim t_0$ of length $t_0$ give the same contributions to $\Delta L$ and $S$, if 
\begin{equation}\label{E068}
p_0\propto \frac{1}{\sqrt{t_0}}\mbox{~~~~or~~~~} \lambda\propto -\frac{1}{t_0^{3/2}}\,.
\end{equation}
As a result, we may assume that if
\begin{equation}\label{E070}
\lambda(t) =  -\frac{p_\lambda t_\lambda^{1/2}}{(t_\lambda-t)^{3/2}}\mbox{~~~~or~~~~}
p(0,t)\sim p_\lambda \sqrt{\frac{t_\lambda}{t_\lambda-t}}
\end{equation} 
at the interval 
\begin{equation}\label{E072}
t_\lambda\ll (-t) \ll 1\, ,
\end{equation} 
when $p_\lambda\lesssim 1$ and $t_\lambda\ll1$, then each octave in $(-t)$ give the same contribution to $\Delta L$ and $S$. This contribution can be estimated in accordance to Eqs.~(\ref{E010}) and~(\ref{E020}) as $\delta\, \Delta L\sim - p_\lambda\sqrt{t_\lambda}$ and $\delta S\sim p_\lambda^2 t_\lambda$, respectively, regardless of $t$ belonging the interval~(\ref{E072}). The number of such octaves can be estimated as $\ln t_\lambda^{-1}$. Hence the total edge displacement and the total action can be estimated as $\Delta L\sim - p_\lambda\sqrt{t_\lambda}\ln t_\lambda^{-1}$ and $S\sim p_\lambda^2 t_\lambda \ln t_\lambda^{-1}$, respectively. Such relationships lead  to $S\sim \Delta L^2 /|\ln \Delta L^2|$. This action is much less than for the trial functions considered  in Sec.~\ref{1sc}, and obeys the condition~(\ref{E004}). Such rough estimation cannot give the correct overall numerical factor of the order of 1 in the latter expression. However this consideration gives some insight into the origin of much smaller actions for multi-scale time trial functions. We may see that power laws in Eq.~(\ref{E070}) are actually quite similar with what we set in Eqs.~(\ref{lam20}) and~(\ref{H126}). 

\paragraph*{\underline{A power law trial function for $\lambda(t)$.}} -- We see that power-law functions for $\lambda(t)$ could lead to the condition~(\ref{E004}). We consider here the following general power  trial functions for $\lambda(t)$:
\begin{equation}\label{E090}
\lambda(t) =  -\frac{p_\lambda t_\lambda^{\alpha-1}}{(t_\lambda-t)^{\alpha}} \, .
\end{equation} 
We consider such  solutions of the OFM equations at the time interval~(\ref{E072}), assuming $t_\lambda \ll 1$ and $p_\lambda\lesssim 1$. We assume that $\lambda(t)$ tends quickly to 0 for $(-t) \gtrsim 1$, and $\lambda(t)\sim - p_\lambda /\sqrt{t_\lambda}$. Then the solution for $p(x,t)$ in the domain $\bar{\Omega}$ becomes in accordance to Eq.~(\ref{H064}) as follows.
\begin{equation}\label{E100}
p(x,t)\bigr|_{\bar{\Omega}}\sim \frac{p_\lambda t_\lambda^{\alpha-1}}{(t_\lambda-t)^{\alpha-1}} \, .
\end{equation} 
The solution for $q(x,t)$ inside the domain $\Omega$ can be treated as previously in the case of $p(0,0)\ll 1$. See Fig.~\ref{sk1}. However we should make obvious re-designations: $L_0-\bar{\ell}_0\to t \dot{L}(t)$, and $t_0\to (-t)$. Then we obtain analogously to Eq.~(\ref{E010}):
\begin{equation}\label{E110}
t \dot{L}(t)\sim p(0,t)\sqrt{-t} \,,
\end{equation} 
or
\begin{equation}\label{E120}
\dot{L}(t)\sim -\frac{p(0,t)}{\sqrt{-t}}\sim -\frac{p_\lambda t_\lambda^{\alpha-1}}{(t_\lambda-t)^{\alpha-1/2}}\, .
\end{equation} 
The calculation of $\dot{S}$ is quite similar to obtaining of Eq.~(\ref{E020}) before multiplying $\dot{S}$ in Eq.~(\ref{E020}) on $t_0$. Thus, we have
\begin{equation}\label{E130}
\dot{S}(t)\sim p(0,t)^2\sim\frac{p_\lambda^2 t_\lambda^{2\alpha-2}}{(t_\lambda-t)^{2\alpha-2}}\, .
\end{equation} 
To get the  total edge displacement, $\Delta L$ and the total action, $S$, determined by the time interval $t\in (-1+t_\lambda,0)$, we should integrate the expressions in Eqs.~(\ref{E120}) and~(\ref{E130}), respectively, over $dt$ on this interval. These contributions to $\Delta L$ and $S$ can be written as:
\begin{eqnarray}
\label{E140}
\Delta L &\sim& - \frac{p_\lambda t_\lambda^{1/2}}{|2\alpha-3|}\left|t_\lambda^{\alpha-3/2}-1\right|\, ,\\
\label{E150}
S&\sim& \frac{p_\lambda^2 t_\lambda}{|2\alpha-3|}\left|t_\lambda^{2\alpha-3}-1\right|\, ,
\end{eqnarray}
when $\alpha\ne 3/2$. For sufficiently small $|\alpha-3/2|$ and $t_\lambda$ contributions to $\Delta L$ and $S$ from this interval become considerably higher than contributions from the regions, when $(-t)\gtrsim 1$ and $\lesssim t_\lambda$. Eliminating $p_\lambda$, we obtain from these equations:
\begin{equation}\label{E160}
S\sim |2\alpha-3|\,  \Delta L^2 \frac{\left|t_\lambda^{2\alpha-3}-1\right|}{\left(t_\lambda^{\alpha-3/2}-1\right)^2}\, .
\end{equation}
Tending $t_\lambda\to 0$ we have $\Delta L\to 0$; and $S$ can be expressed in this limit as
\begin{equation}\label{E170}
S\sim |2\alpha-3|\,  \Delta L^2
\end{equation}
for sufficiently small $|2\alpha-3|$.

We may conclude that the lowest action will take place at $\alpha\to3/2$. For any finite $|\alpha-3/2|$ and sufficiently small $t_\lambda$ we may make $S$ lower at $\Delta L\to0^-$ by choosing lower $|\alpha-3/2|$. It means that $\alpha = 3/2$ corresponds to the optimal $\lambda(t)$ in the form of Eq.~(\ref{E090}), if we consider the leading-order behavior of $S$ at $\Delta L\to 0^-$. Namely this set of $\lambda(t)$ was considered analytically and numerically in Sec.~\ref{ps}.

\paragraph*{\underline{Power-law trial function 
 with slowly varying amplitude.}} -- It is interesting to introduce in Eq.~(\ref{E090}) a very slowly variable factor at $\alpha=3/2$, trying to diminish the leading order in the expressions~(\ref{SS220}) for the action. We assume that the change of the factor  is relatively small if we multiply or divide the time $t$ by 2. As a result, we present $\lambda(t)$ in the form:
\begin{equation}\label{E190}
\lambda(t) =  -\frac{ t_\lambda^{1/2}}{(t_\lambda-t)^{3/2}} F\left(\ln\frac{1}{|t|},\ln\frac{1}{t_\lambda}\right)
\end{equation} 
We assume again that this expression is valid for $t_\lambda\ll (-t)\ll 1$. Contributions to action outside this interval again determines only subleading orders at $\Delta L\to 0^-$. When absolute value of partial derivative of the function $F$ with respect to the 1st argument is much less than 1, then dependence of $F$ on $t$ can be treated adiabatically. Then repeating previous estimations we can write:
\begin{equation}\label{E200}
 \Delta L\sim - \sqrt{t_\lambda}\int_{t_\lambda}^1 F\left(\ln\frac{1}{|t|},\ln\frac{1}{t_\lambda}\right)\, \frac{d|t|}{|t|}\, ,
\end{equation}
and
\begin{equation}\label{E210}
 S\sim t_\lambda\int_{t_\lambda}^1 F^2\left(\ln\frac{1}{|t|},\ln\frac{1}{t_\lambda}\right)\, \frac{d|t|}{|t|}\, .
\end{equation}
Hence
\begin{equation}\label{E220}
 S\sim\Delta L^2\frac{\int_{t_\lambda}^1 F^2\left(\ln\frac{1}{|t|},\ln\frac{1}{\Delta\ell^2}\right)\, \frac{d|t|}{|t|}}{\left[\int_{t_\lambda}^1 F\left(\ln\frac{1}{|t|},\ln\frac{1}{\Delta\ell^2}\right)\, \frac{d|t|}{|t|}\right]^2}\, .
\end{equation}
Minimizing this expression at given $\Delta L$, we obtain that optimal $F$ has not to depend on $\ln |t|$: 
\begin{equation}\label{E230}
F=\mbox{const}=p_\lambda\, .
\end{equation}

As a result we may conclude that optimal $\lambda(t)$ has to have a form of Eq.~(\ref{E070}) at $\Delta L\to 0^-$. The only question that should be solved is the question about the amplitude $p_\lambda$ in Eq.~(\ref{E070}).

\paragraph*{\underline{A choice of the constant $p_\lambda$ in Eq.~(\ref{E070}).}} -- As a consequence of the arguments above, $p_\lambda$ for the optimal $\lambda(t)$ cannot be much larger than 1. Thus we set straightly that $p_\lambda\lesssim 1$ in the optimum. For such $p_\lambda$ we are able to make substitution $p_\lambda^2 t_\lambda$ instead of $t_\lambda$ in the amplitude of $\lambda(t)$ in the definition~(\ref{H126}) of Sec.~\ref{aps}, where we considered analytic solution of the OFM equations with $\lambda(t)$ defined in Eq.~(\ref{H126}). Analogous substitutions in all further expressions in that section lead to the following slightly more general final results than in Sec.~\ref{aps} [Eqs.~(\ref{SS212}) and~(\ref{SS214})]:
\begin{equation}\label{G010}
\Delta L=p_\lambda \sqrt{\frac{t_\lambda}{\pi}}\, \Bigl[ \ln  t_\lambda+{\cal O}(1)\Bigr]\, .
\end{equation}
\begin{equation}\label{G020}
S=\frac{p_\lambda^2 t_\lambda}{4}\,  \left[\ln \frac{1}{t_\lambda}+{\cal O}(1)\right]\, .
\end{equation}
It is worth to remind once again that the residual terms, ${\cal O}$ in these equations are of the order of 1; and they are determined by by unknown behavior of $\lambda(t)$ at $(-t)\gtrsim 1$ and $\sim t_\lambda$ in the optimum. These equations give:
\begin{equation}\label{G030}
S(L) =\frac{\pi}{4} \frac{\Delta L^2}{\ln \left(p_\lambda/\Delta L\right)^2}+\dots\, 
\qquad(\Delta L\to 0^-)\,.
\end{equation}
If $p_\lambda\sim 1$, then it can be skipped at all or transferred to the residual term. However, when $p_\lambda \ll 1$, it leads to an increase of the trial action. This means that the optimal $p_\lambda\sim 1$ and its exact value do not influence on the asymptotic behavior of $S$ in the leading order. Our choice $p_\lambda=1/4$ in Sec.~\ref{ps} follows this conclusion; and its concrete numerical value was chosen only for numerical convenience.

\paragraph*{\underline{Final OFM result.}} -- Combining now the OFM results~(\ref{G010}) and~(\ref{G020}),  we obtain similarly to obtaining of Eq.~(\ref{SS230}):
\begin{eqnarray}
\label{G050}
S(L) &=&\frac{\pi}{4} \frac{\Delta L^2}{|\ln\Delta L^2|}\, \biggl[1-2\frac{\ln |\ln\Delta L^2|}{|\ln\Delta L^2|} \nonumber\\[1mm]
&+&{\cal O}\left(\frac{1}{|\ln\Delta L^2|}\right)\biggr]
\qquad(\Delta L\to 0^-)\,.
\end{eqnarray}
Our main statement is that this is the valid OFM result at $\Delta L\to 0^-$. The non-optimized value of the overall factor $\sim{\cal O}(1)$ in our trial function $\lambda(t)$, as well as its non optimized behavior at $(-t)\gtrsim 1$ and at $(-t)\sim t_\lambda$ may change only the $\sim{\cal O}(1)$ coefficient before $\ln^{-1}(\Delta L)^{-2}$ in the residual term ${\cal O}\left(\ln^{-1}(\Delta L)^{-2}\right)$ that is of the order of 1. The order of this residual term is confirmed by the numerical simulations in Sec.~\ref{nps}.

The result~\cite{Si21} concerning variance of $L$ for typical fluctuations means that $P(L)$ can be presented for such fluctuations as
\begin{equation}\label{G060}
-N^{-1}\, \ln P(L) \simeq \frac{\pi}{4} \frac{\Delta L^2}{\ln N}\,.
\end{equation}
Our main result~(\ref{I040}) followed from Eq.~(\ref{G050}) coincides with Eq.~(\ref{G060}) with the relative accuracy $\varepsilon\to 0$, when $\Delta L$ belongs for example the interval ${\cal L}=(-1/\sqrt{N^{1-\varepsilon}}, -1/\sqrt{N})$. For any small $\varepsilon$ and sufficiently high $N$, $|\Delta L|$ varies on this interval $\cal L$ in many times. It means that the result (\ref{G060}) for typical fluctuations and our result~(\ref{I040}) have a wide region near the point $\Delta L=\sqrt{\mbox{var}\,(L)}$, where they coincide with a high accuracy. This fact strengths reliability of the both results.

\section{Positive atypical fluctuations of $L$: Single particle approximation}
\label{SP}

\begin{figure}[ht]
  \centering
  \includegraphics[width=8cm]{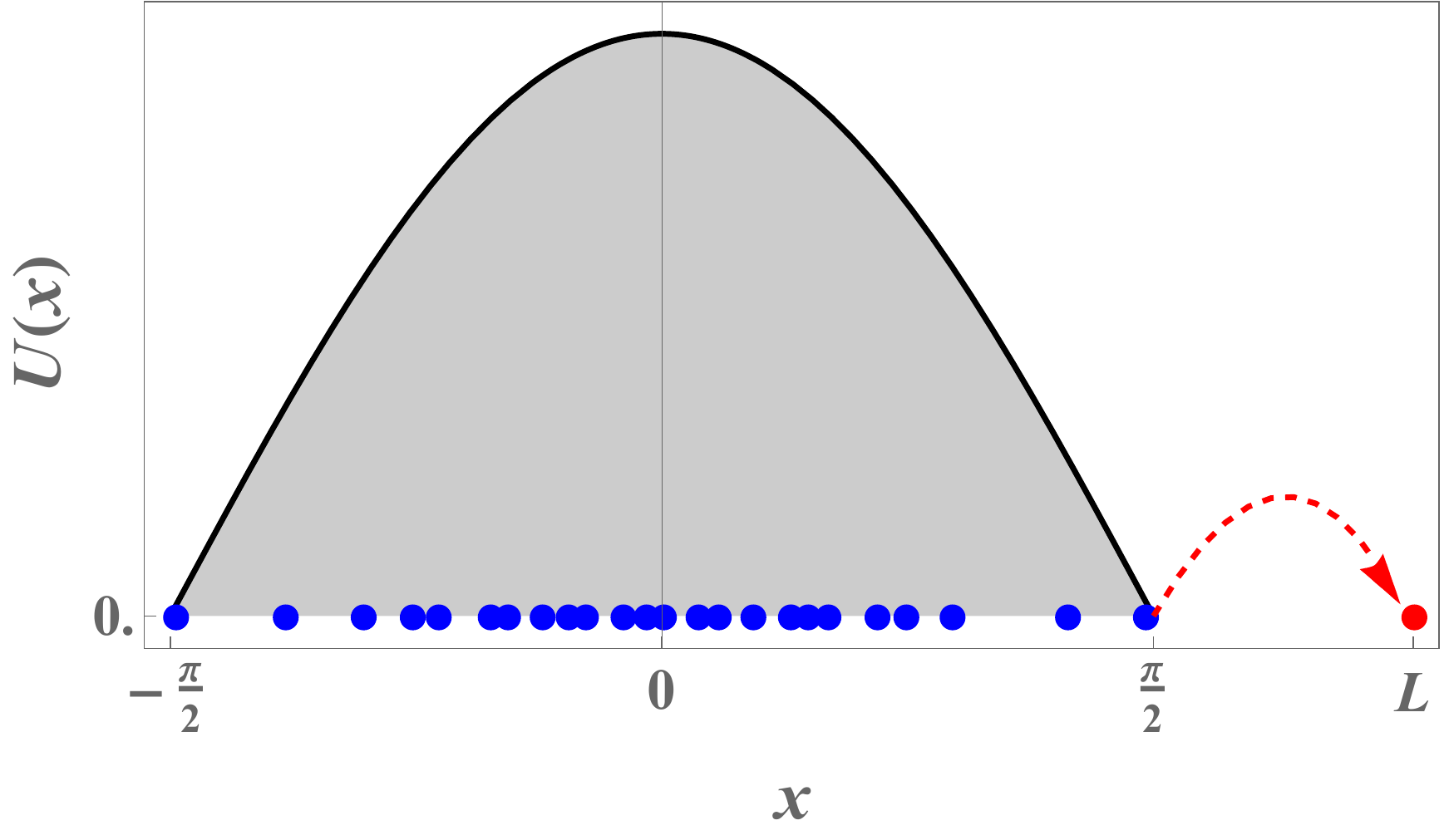}
  \caption{Atypical positive fluctuations of $\Delta L$ are dominated by the ``runaway particle'' scenario, in which a single particle quickly travels from the edge support of $U(x)$, at $x = \pi/2$, to the position $x=L$, while the rest of the particles do not display any unusual behavior. During the creation of this fluctuation, the branching process is entirely suppressed.}
  \label{figRightTail}
\end{figure}

Fluctuations with an unusually-large swarm radius, $L > \bar{\ell}_0$, turn out to behave entirely differently to the case $L < \bar{\ell}_0$ that we considered in the previous section.
As we find below, the system trajectories that dominate the probability for observing some value $L > \bar{\ell}_0$ are those for which a single, runaway particle travels relatively quickly from $x = \bar{\ell}_0$ to $x=L$, whereas the other particles simply diffuse, and meanwhile the branching process is completely suppressed. This scenario is schematically depicted in Fig.~\ref{figRightTail}.

The runaway particle scenario is similar in spirit to similar approaches in extreme-value statistics \cite{MPS20}. In particular, it is rather reminiscent of the ``evaporation'' scenario that describes the right tail of the statistics of the largest eigenvalue in many random matrix ensembles \cite{KTJ76, JKT78, FMPPS08, NMV10, FFPPY13}. In fact, this approach was also recently employed to describe fluctuations of the size of a model that is not so different to the Brownian bees model, in which the branching process is replaced by stochastic resetting of the particles' positions to the origin \cite{VAM22}.

Assuming this scenario of a single, runaway particle, the problem simplifies considerably.
One can write down a very simple equation for the dynamics of the  PDF $P_{1}\left(x,t\right)$ of the particle that is furthest from the origin, by neglecting the possibility that it will be overtaken by one of the other particles:
\begin{equation}\label{SI010}
\partial_t P_1(x,t)=\partial_x^2 P_1(x,t) - N P_1(x,t)\, ;\quad (x-\bar{\ell}_0  > 0)\,.
\end{equation}
The two terms on the right hand side of Eq.~\eqref{SI010} describe diffusion of the furthest particle from the origin, and an effective `mortality' term that corresponds to branching of one of the other $N-1$ particles (we neglect here the difference between $N$ and $N-1$ at $N \gg 1$).
This equation is expected to be valid at $x-\bar{\ell}_0$ that is much larger than the scale of typical fluctuations, where it becomes very unlikely for the positions of the two furthest particles from the origin to cross each other.
The steady state solution of this equation is quite obvious $P_1(x)\propto e^{-x\sqrt{N}}$. As a result we find that positive fluctuations of $L$ are described by
\begin{equation}\label{SI020}
P(L)= P_1(L)\sim \exp(-\sqrt{N}\Delta L)\quad (\Delta L > 0)\,.
\end{equation}
A prefactor in this equation is determined by a crossover region, where one expects the PDF~(\ref{G060}) for typical fluctuations to match somehow with Eq.~(\ref{SI020}). A full calculation of the prefactor is beyond the quantitative theory presented here. However, one may assume that this crossover takes place at $\Delta L$ for which the two formulas (\ref{G060}) and (\ref{SI020}) predict probabilities that are of the same order. This happens at $\Delta L\sim \sqrt{\ln (N)\, \mbox{var}\, (L)}$. In any case, Eq.~(\ref{SI020}) gives Eq.~(\ref{I050}) from the Introduction.

Some insight is obtained by contrasting the results of this section with those of the previous one, which describe the two (very different) tails of the distribution $P(L)$.
The scaling $-\ln P(L) \sim \sqrt{N}$ predicted by the runaway particle scenario of the present section, obviously predicts much larger probabilities than the scaling $-\ln P(L) \sim \sqrt{N}$ predicted by the OFM of the previous section, see e.g. \eqref{H050}. This confirms our assumption that, for $\Delta L > 0$, the runaway particle scenario dominates, whereas scenarios involving a large number of particles should not be taken into account as their contribution to $P(L)$ is negligible.

However, in analogy with the previous section, it would be nice to gain further information regarding the  atypical $\Delta L > 0$ fluctuations, by characterizing the histories of the system that lead to a given $L > \bar{\ell}_0$. It turns out that this can be done quite simply, as follows.
Let us consider a dynamical scenario in which, at time $t= - \tau$ (where $\tau \gg 1/N$ will be determined below) the system is in a state that is described by the density $U(x)$. Then, during the time interval $-\tau < t < 0$, (i) no branching events occur, and (ii) the rightmost particle travels from the edge of the support of $U(x)$, $x=\bar{\ell}_0$, arriving at $x=L$ at time $t=0$. This scenario is described schematically in Fig.~\ref{figRightTail}.

What is the probability of this dynamical scenario? The probability for no branching events is (exactly) given by $e^{-N\tau}$. Conditioned on no branching events, the PDF of the position a particle initially at time $t=0$ given that at time $t=-\tau$ it was at $x=\bar{\ell}_0$ is
$e^{-\left(x-\bar{\ell}_{0}\right)^{2}/4\tau}/\sqrt{4\pi \tau}$.
Therefore, the probability for this scenario, including arrival at $x=L$ at time $t=0$, is
\begin{equation}
\sim e^{-N\tau-\left(\Delta L\right)^{2}/4\tau} \, ,
\end{equation}
up to a pre-exponential factor. It will be useful to rewrite this as
\begin{equation}
\label{probRunawaytau}
\sim e^{-\sqrt{N}\mathcal{F}\left(\tilde{\tau}\right)}\,,\quad\mathcal{F}\left(\tilde{\tau}\right)=\tilde{\tau}+\frac{\left(\Delta L\right)^{2}}{4\tilde{\tau}},\quad\tilde{\tau}=\sqrt{N}\,\tau \, .
\end{equation}
The next step towards calculating the $\Delta L > 0$ tail of $P(L)$ is to integrate the probability \eqref{probRunawaytau} over $\tilde{\tau}$. Clearly, at $N \gg 1$ this integral is dominated by the saddle point, i.e., we obtain
\begin{equation}
\label{probRunawayF}
P\left(L\right)\sim e^{-\sqrt{N}\mathcal{F}\left(\tilde{\tau}_{*}\right)}
\end{equation}
where $\tilde{\tau}_{*}$ is the minimizer of $\mathcal{F}\left(\tilde{\tau}\right)$.
This minimization is trivial; it yields $\tilde{\tau}=\Delta L/2$ so $\mathcal{F}\left(\tilde{\tau}\right) = \Delta L$, which, after plugging into \eqref{probRunawayF}, we obtain $P\left(L\right)\sim e^{-\sqrt{N}\Delta L}$ in perfect agreement with our earlier result \eqref{SI020}.

\section{Summary and Discussion}
\label{summ}

We see that the PDF for the size $L$ of the swarm in the frame of the `Branching Bees' model with $N\gg 1$ is quite asymmetric around its mean value $\bar{\ell}_0$, if we exclude at least the region of typical fluctuations of $L$ determined by its variance~(\ref{L140})~\cite{Si21}. 
In particular, we find that unusually large positive fluctuations of $L$ are far more likely than negative ones, as is evident from the very different scalings of the two distribution tails with $N$ at $N \gg 1$.
The atypically large negative fluctuations of $L$ can be described by the OFM approach~(\ref{I040}); and this PDF demonstrates the logarithmic anomaly that also appears in the variance ~(\ref{L140})~\cite{Si21}. The OFM result matches smoothly with the Gaussian PDF determined by the variance. For atypically large positive fluctuations of $L$, their PDF~(\ref{I050}) can be obtained with a single runaway particle approach. 
The region of crossover of the PDF between the latter one behavior and the Gaussian part of the PDF for typical fluctuations is an interesting goal for further investigations.

We saw that for $|\Delta L| \ll 1$ the fluctuations involve mainly a narrow layer of bees close to $\bar{\ell}_0$. As a result, we may assume that the principal results of this paper concerning $P(L)$ at $|\Delta L|\ll1$ do not depend on dimension $d$ of the space (up to a proper shift of the distribution $P(L)$, because $\bar{\ell}_0$ depends on $d$). The Monte-Carlo simulations in Ref.~\cite{Si21} for typical fluctuations and conclusions from them drawn there support this argument.

A model, which is similar but slightly simpler than the Brownian bees model, was recently considered in Ref.~\cite{VAM22}. In their `model B', the position of the particle farthest from the origin is stochastically reset to the origin (instead of being reset to the position of one of the other particles as in the Brownian bees model studied here).
We believe that our main results, Eqs.~(\ref{I040}) and~(\ref{I050}), remain valid for the `model B' also  (up to a proper shift, again because $\bar{\ell}_0$ is different). We draw such conclusion for negative $\Delta L$ from the fact that we were able to neglect the branching process for the self similar solution in the domain $\Omega$, see Eq.~(\ref{SS060}). For the positive $\Delta L$ this conclusion is even more obvious and derived actually in Ref.~\cite{VAM22}. Moreover, we may assume that the entire PDF of $L$ at $|\delta L|\ll 1$ and $N\to\infty$ for these two models are the same (up to the shift). Meanwhile, this statement is proven for the variance, $\mbox{var}(L)$, in Ref.~\cite{VAM22}.

\begin{acknowledgments}

We are very grateful to Baruch Meerson for useful discussions. 
The research of P.S. was supported by the project High Field Initiative (CZ.02.1.01/0.0/0.0/15\_003/0000449)
from the European Regional Development Fund.
\end{acknowledgments}

\appendix
\section{Numerical method}
\label{AppendixNumerical}

We describe here some details of the numerical analysis of the problem~(\ref{H062}), (\ref{H064}), (\ref{H030}), (\ref{H040}), (\ref{H080}) and~(\ref{conservq}), in one dimension,
where the constant $\Lambda$ and functions $L(t),\quad q(x,t),\quad p(x,t)$ should be found, and $\lambda(t)$  is a given function, see  Eq.~(\ref{lam20}).
We introduce a new spatial variable $y=x/L(t)$ ($|y|\leq 1$) to work with a stationary spatial grid.
This change of variables causes Eqs.~(\ref{H062}), (\ref{H064}) to become
\begin{eqnarray}\label{qyt}
\!\!\!\!\!\!\!\! \partial_{t}q &=& y\frac{\dot{L}(t)}{L(t)}\partial_{y}q+\partial_{y}^{2}q-\partial_{y}(\partial_{y}q-2q\partial_{y}p)+q e^{p}\,,\\
\label{pyt}
\!\!\!\!\!\!\!\! \partial_{t}p&=&y\frac{\dot{L}(t)}{L(t)}\partial_{y}p-\partial_{y}^{2}p-(\partial_{y}p)^{2}-(e^{p}-1)-\lambda(t)
\end{eqnarray}
respectively,
while the initial and boundary conditions become
\begin{eqnarray}\label{bc1}
&&q(|y|=1,t)=p(|y|=1,t)=0\,,\\
\label{bctinf}
&&q(y,t\rightarrow-\infty)=\frac{1}{2}\cos\left(\frac{\pi}{2}y\right);\quad p\left(y,t\rightarrow-\infty\right)=0\,,\nonumber\\\\
\label{pt0}
&&p(y,0)=\Lambda=\mbox{const},\quad \mbox{for}\quad |y|<1\,.
\end{eqnarray}
In our numerical solutions, we replace time $\infty$ by finite time $T=10$. The conservation condition  \eqref{conservq} is:
\begin{equation}\label{qintgrl}
L(t)\int_{-1}^{1}q(y,t)dy=1\,.
\end{equation} 

As described in the main text, we solve Eqs.~\eqref{qyt} and \eqref{pyt} using the back-and-forth Chernykh-Stepanov algorithm \cite{ChSt}.
Every iteration of the algorithm consists of two steps. In the first step, we solve Eq.~(\ref{pyt}) for $p(y,t)$ backwards in time from $t=0$, using $q(y,t)$ and $L(t)$ from the previous iteration. In the second step, we solve Eq.~(\ref{qyt}) forward in time for $q(y,t)$ using $p(y,t)$ that was found in the first step. During the forward step, we also compute $L(t)$ via Eq.~\eqref{qintgrl}.
We employ the implicit finite differences method to approximate Eqs.~\eqref{qyt} and \eqref{pyt}, and Newton's method to solve nonlinear algebraic equations to approximate $p$ . 
A few iterations of the algorithm are sufficient for it to achieve convergence to a solution of Eqs.~\eqref{qyt} and \eqref{pyt}.
One must, of course, specify some initial values in order to perform the first iteration of the algorithm. The choice usually does not have a strong effect on the convergence properties.
We chose
$\Lambda=\Lambda_{1}\cong 1/2$, the mean field functions $q(0<y<1,-T\leq t<0)=\cos (L(t) y)/2$ and $p(y,t<0)=0,\quad p(y,0)=\Lambda_{1}$ and  $L(t)=\bar{\ell}_0$.

At the end of the iterations  the functions $q(y,t,\Lambda_{1}),\quad p(y,t,\Lambda_{1})$ and $L(t,\Lambda_{1})$, in particular $p(y,-T,\Lambda_{1})$ are known. The found function $p(y,-T,\Lambda_{1})$ satisfies to condition  $p(y,-\infty)=0$ in Eq.~(\ref{bctinf}) for a certain value of $\Lambda$ which we are to find. Since in the general case $p(y,-T,\Lambda_{1})$ is not a constant, we employ a functional $F(\Lambda)=\int_{-1}^{1}p^{2}(y,-T)dy$. We seek now the value $\Lambda_{m}$ that minimizes the functional $F$, which is calculated by the procedure described above. Notice, that $F(\Lambda_{m})$ is very close to zero for sufficiently small criteria ending the iterations.
Finally $L$ and $S$ are computed on the solution through $L = L(t=0)$ and Eq.~\eqref{H090}, respectively.

We used non homogeneous time and space grids. The smallest time step $\tau_{m} $ at $t=0$ depends on value of $t_{\lambda}$, $\tau_{m}\propto t_{\lambda}^2$ the value of the steps growth exponentially when $t \rightarrow -T$: $\tau_{j}=(1+\delta\tau)\tau_{j+1}$, $t_{j}=t_{j-1}+\tau_{j}$. $j=2,3,...,m$, $t_{1}=-T$ and $t_{m}=0$. For example $\tau_{m}=10^{-13}$ and  $\delta\tau=0.01$ in calculation of version for $t_{\lambda}=10^{-6}$. The space grid is exponential too, the minimal spacing is  near $y=1$: $h_{2}\propto t_{\lambda}^{1/2}$, and the spacing increases with growth of $y$. For  $t_{\lambda}=10^{-6}$, we used $h_{2}=5\cdot 10^{-5}$ and  the increment $\delta h=0.01$, $1-y_{i+1}=(1+\delta h)(1-y_{i})$, $i=1,2,3,...,n$, $y_{1}=1,\quad y_{n}=0$. 

To illustrate the calculation of $\Lambda$, we give here the data when the shooting procedure was stopped  for $t_{\lambda}=10^{-6}$:  $p(0,-T)\simeq -10^{-12}$, $\int_{-1}^{0} |p(y,-T)|dy\simeq 3\times10^{-12}$, $\int_{-1}^{0}p(y,-T) \cos(y) dy\simeq-3\times 10^{-12}$. 

It is important also that our numerical model has an steady-state mean field distribution which differs because of its discreetness  from the continuous distribution $U(x)$, defined by Eq.~(\ref{Ux}). This difference really is very small. Nevertheless, we took it into account in the calculation of numerical $\Delta L$ to diminish the influence of numerical inaccuracies.

\end{document}